\documentclass[useAMS,usenatbib]{mn2e}
\usepackage{color}
\usepackage[pdftex]{graphicx}

\usepackage{multirow}
\usepackage{amssymb}

\usepackage{amsmath}

\newcommand{\pc}{\mbox{$\>{\rm pc}$}} 
\newcommand{\kpc}{\mbox{$\>{\rm kpc}$}} 
\newcommand{\Gyr}{\mbox{$\>{\rm Gyr}$}}
\newcommand{\Myr}{\mbox{$\>{\rm Myr}$}}

\newcommand{\Msun}{\>{\rm M_{\odot}}}

\newcommand\degrees{^\circ}
\newcommand{\avg}[1]{\mbox{$\left<{#1}\right>$}}

\newcommand{\feh}{\mbox{$\rm [Fe/H]$}}

\newcommand{\tform}{\mbox{$t_\mathrm{f}$}}

\newcommand{\thesim}{\textsf{m12m}}

\def\ie{{\it i.e.}}

\title[Cosmological X-shaped bulge]{Formation, vertex deviation and age of the
  Milky Way's bulge: input from a cosmological simulation with a
  late-forming bar}
 
\author[Debattista et al.]{Victor P. Debattista$^1$\thanks{E-mail:
    vpdebattista@gmail.com}, Oscar A. Gonzalez$^2$, Robyn
  E. Sanderson$^3$, \newauthor Kareem El-Badry$^4$, Shea
  Garrison-Kimmel$^3$\thanks{Einstein Fellow}, Andrew Wetzel$^5$,
  \newauthor Claude-Andr\'e Faucher-Gigu\`ere$^6$,
  Philip F. Hopkins$^3$ \\
$^1$ Jeremiah Horrocks Institute, University of Central Lancashire,
  Preston, PR1 2HE, UK \\
$^2$ UK Astronomy Technology Centre, Royal Observatory, Blackford
  Hill, Edinburgh EH9 3HJ, UK \\
$^3$ TAPIR, MC 350-17, California Institute of Technology, Pasadena,
  CA 91125, USA \\
$^4$ Department of Astronomy and Theoretical Astrophysics Center,
  University of California Berkeley, Berkeley, CA 94720 \\
$^5$ Department of Physics, University of California, Davis, CA 95616,
  USA \\
  $^6$ Department of Physics and Astronomy and Center for
  Interdisciplinary Exploration and Research in Astrophysics (CIERA), \\
  Northwestern University, 2145 Sheridan Road, Evanston, IL 60208, USA \\ }

\begin{document}   

\date{{\it Draft version on \today}}
\pagerange{\pageref{firstpage}--\pageref{lastpage}} \pubyear{----}
\maketitle

\label{firstpage}

\begin{abstract}
We present the late-time evolution of \thesim, a cosmological
simulation of a Milky Way-like galaxy from the FIRE project. The
simulation forms a bar after redshift $z=0.2$.  We show that the
evolution of the model exhibits behaviours typical of {\it kinematic
  fractionation}, with a bar weaker in older populations, an X-shape
traced by the younger, metal-rich populations and a prominent X-shape
in the edge-on mean metallicity map.  Because of the late formation of
the bar in \thesim, stars forming after $10\Gyr$ ($z=0.34$)
significantly contaminate the bulge, at a level higher than is
observed at high latitudes in the Milky Way, implying that its bar
cannot have formed as late as in \thesim.
We also study the model's vertex deviation of the velocity ellipsoid
as a function of stellar metallicity and age in the equivalent of
Baade's Window.  The formation of the bar leads to a non-zero vertex
deviation.  We find that metal-rich stars have a large vertex
deviation ($\sim 40\degrees$), which becomes negligible for metal-poor
stars, a trend also found in the Milky Way, despite not matching in
detail.  We demonstrate that the vertex deviation also varies with
stellar age and is large for stars as old as $9 \Gyr$, while $13\Gyr$
old stars have negligible vertex deviation.  When we exclude stars
that have been accreted, the vertex deviation is not significantly
changed, demonstrating that the observed variation of vertex deviation
with metallicity is not necessarily due to an accreted population.
\end{abstract}



\section{Introduction}
\label{sec:intro}

Our understanding of the formation of the bulge of the Milky Way (MW)
has advanced considerably, with large new observational surveys
\citep[e.g.][]{howard+08, argos, vvv, zoccali+14, apogee, ness+12}, careful
comparison with simulations \citep[e.g.][]{jshen+10,
martinez-valpuesta_gerhard11, martinez-valpuesta_gerhard13,
pdimatteo+15, pdimatteo16, debattista+17, athanassoula+17,
fragkoudi+17b, fragkoudi+18, buck+18, buck+18b}, and detailed
dynamical models of its current state \citep[e.g.][]{bissantz+04,
portail+15, portail+17}.  All three approaches have now deconstructed
the bulge by stellar populations, demonstrating how its properties
vary as a function of metallicity \citep{ness+13a, pdimatteo+15,
debattista+17, portail+17, athanassoula+17, fragkoudi+18}.
Multiple studies have converged to the conclusion that the majority of
the bulge formed purely from the secular evolution of the disc, via
the bar that forms within it.  Based on the kinematics of M-giants
observed in BRAVA \citep{howard+08}, \citet{jshen+10} estimated that
any accreted component constitutes less than $8\%$ of the stellar mass
of the MW, while \citet{debattista+17} showed that the presence of a
hot component only becomes evident at low metallicities, where the
addition of $1.3\%$ of the total stellar mass in kinematically hot
stars is sufficient to match the kinematics of these stars.  In a
similar vein \citet{pdimatteo+14} estimated that a classical bulge
with $25\%$ of the disc mass can be excluded. \citet{bonaca+17} and
\citet{elbadry+18b} showed that the kinematics of old, accreted,
metal-poor stars in the central spheroid are indistinguishable from
those of the stars of the same age that formed in situ.  Therefore the
observed population of kinematically hot stars must include stars that
formed in situ, making the contribution of an accreted population even
lower.  Properties that the secular evolution model can now account
for include the vertical metallicity gradient, the predominantly old
stars in the bulge, the age and metallicity variation of the X-shape
and bar strength, and the different kinematics of stars of different
age.  The key mechanism driving the observed trends with stellar
populations is the separation of stellar populations by an evolving
bar on the basis of their radial velocity dispersions, a process
termed {\it kinematic fractionation} by
\citet{debattista+17}.  This occurs because kinematically hot
populations have a lower angular frequency relative to the bar.  The
frequency at which they encounter a vertical bend in the bar is
therefore lower than for a cool population, allowing them to be pumped
by the bar to larger heights before their response to the forcing is
out of phase \citep{merritt_sellwood94}.  Since stellar populations
typically get kinematically hotter as they age, kinematic
fractionation generally results in a continuum of properties as a
function of age.  Starting with \citet{bekki_tsujimoto11}, and
subsequently \citet{pdimatteo16} and \citet{fragkoudi+17} reached a
similar conclusion using simulations composed of distinct thin and
thick discs.  While stars in the simulation of \citet{debattista+17}
all form self-consistently from gas, the simulation was evolved in
isolation, removed from a larger scale cosmological context.  Recently
\citet{buck+18} demonstrated that the signatures of kinematic
fractionation also occur in a cosmological simulation.  Here we
confirm this result using a cosmological simulation, \thesim, from the
Feedback In Realistic Environments (FIRE) project.

One of the properties of the bulge which is yet to be explained
without invoking an accreted population is the absence of a
significant vertex deviation in the most metal-poor stars of the bulge
\citep{soto+07, babusiaux+10}.  The vertex deviation measures the
covariance between radial and tangential motions (from the Sun's point
of view).  A stationary, axisymmetric disc has no vertex deviation,
whereas a triaxial bar necessarily introduces a vertex deviation
\citep{binney_tremaine08}.  Observations show that the metal-rich
stars in Baade's Window ($(l,b) = (1\degrees,-4\degrees)$) have a
significant vertex deviation, while the metal-poor stars do not.  This
has often been interpreted as the signature of a separate component in
the bulge \citep[e.g.][]{noguchi99, aguerri+01}.  However
\citet{debattista+17} showed that, in their simulation which did not
have {\it any} accreted population, the oldest population hosts a
substantially weaker bar than the rest of the stars.  Here we explore
whether the vanishing vertex deviation of old stars depends upon the
formation location (in-situ versus accreted).

A further uncertainty about the MW's bar is its age.  Since a bar is
formed from stars in the disc, a bar will always contain stars older
than itself.  But the bar also grows over time, by shedding angular
momentum \citep[e.g.][]{weinberg85, debattista_sellwood00,
athanassoula02, oneill_dubinski03, martinez-valpuesta+06}, with the
possibility of trapping stars that are younger than the bar itself
\citep[e.g.][]{aumer_schoenrich15}.  Therefore measuring the age of
the MW's bar is difficult.  Studies of the age distribution of stars
in the bulge have generally found old stars \citep{ortolani+95,
  kuijken_rich02, zoccali+03, sahu+06, clarkson+08, clarkson+11,
  brown+10, valenti+13, calamida+14}.  In contrast, spectroscopy of
microlensed dwarfs has found a wide range of stellar ages in the
bulge, including very young stars at high metallicity
\citep{bensby+11, bensby+13, bensby+17}.  More recently,
\citet{haywood+16} have proposed that the bulge hosts stars between
$13\Gyr$ and $3\Gyr$ old to explain the narrow range of turnoffs.
\citet{bernard+18} found that over $80\%$ of stars on the bar's minor
axis are older than $8\Gyr$ but that a significant fraction of
super-solar metallicity stars are younger and that $11\%$ of all stars
on the minor axis are younger than $5\Gyr$. All these studies agree
that young stars are predominantly or exclusively found at high
metallicities and, therefore, not expected to be found at high
Galactic latitudes, where low metallicity stars dominate
\citep[e.g.][]{zoccali+17}. These studies however have not provided
constraints on the age of the bar. Alternatively, \citet{buck+18}
propose that the variation of the X-shape as a function of age can be
used to determine the age of the bar.  Here we show what the
consequences for stellar populations on the minor axis would be if the
bar is as young as $2-3\Gyr$.

This paper is organised as follows.  Section \ref{s:simulation}
describes the simulation we use.  This is followed in Section
\ref{s:kfractionation} by several lines of evidence that the bar in
this simulation drives kinematic fractionation.  Section
\ref{s:vertexdeviation} examines the vertex deviation of the model, to
test whether in-situ populations can have negligible vertex deviation.
Section \ref{s:agemwbar} derives constraints on the age of the MW's
bar.  We conclude in Section \ref{s:discussion}.


\section{Simulation}
\label{s:simulation}

\begin{figure*}
\begin{center}
\includegraphics[width=0.8\textwidth]{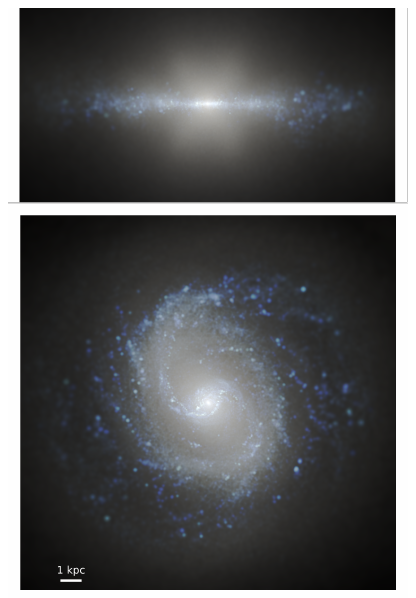}
\caption{Edge-on (top) and face-on (bottom) views of \thesim\ at $z=0$. 
Each image is a u/g/r composite (in \emph{Hubble Space Telescope}
bands) with a logarithmic stretch, using STARBURST99
\citep{leitherer+99} to determine the spectral energy distribution of
each star particle based on its age and metallicity and ray-tracing
following \citet{hopkins+05}. To help reveal the structure of the bar,
no dust extinction is included. The face-on view shows the central
bar, while the edge-on view exhibits a clear X-shape.}
\label{fig:sed-images}
\end{center}
\end{figure*}

The simulation analyzed in this paper, referred to as \thesim, is part
of the Feedback In Realistic Environments
(FIRE)\footnote{{fire.northwestern.edu}} project, specifically the
``FIRE-2'' version of the code; all details of the methods are
described in \citet{Hopkins2018fire2}, Section~2. The simulations use
the code GIZMO
\citep{Hopkins2015gizmo}\footnote{{tapir.caltech.edu/$\sim$phopkins/Site/GIZMO.html}},
with hydrodynamics solved using the mesh-free Lagrangian Godunov
``MFM'' method. Both hydrodynamic and gravitational (force-softening)
spatial resolution are set in a fully-adaptive Lagrangian manner for
gas (but not for stars and dark matter). The simulation includes
cooling and heating from a meta-galactic background and local stellar
sources from $T\sim10-10^{10}\,$K, star formation in locally
self-gravitating, dense molecular gas, and stellar feedback from
stars, including stellar winds from O, B and AGB stars, SNe Ia and II,
and multi-wavelength photo-heating and radiation pressure, with inputs
taken directly from stellar evolution models. The FIRE physics, source
code, and all numerical parameters are identical to those described in
\citet{Hopkins2018fire2}. The basic characteristics of \thesim\ are
given in Table \ref{tbl:simdetails}. Of interest for this work is
that, like the MW, this simulated galaxy has a strong bar and X-shaped
bulge at redshift $z=0$ (Figure \ref{fig:sed-images}). A movie showing
the time-evolution of \thesim\ from $z\sim 8$ to the present
day\footnote{{tapir.caltech.edu/$\sim$sheagk/movies/stars/m12m\_ref13\_star.mp4}}
shows that although \thesim\ has a turbulent early merger history,
including a nearly equal-mass merger at $z \sim 1.5$ ($t \sim 4.3
\Gyr$), it is relatively peaceful at late times, with only minor
mergers since at least $z \sim 0.5$.  While \thesim\ has a satellite
mass function similar to that of M31 \citep{garrison-kimmel+18b},
interactions are not necessarily the cause of bar formation
\citep[see also][]{zana+18}.

To analyze the structure of \thesim, the simulation was first centered
on the host galaxy by iteratively calculating the \emph{stellar}
center of mass. The galaxy is then aligned by calculating the moment
of inertia tensor for all stars within 20 kpc of the center, and
rotated so that the principal axes of this tensor lie along the three
Cartesian axes, with the $X$ direction pointing along the longest axis
and the $Z$ direction pointing along the shortest axis. Since
\thesim\ has a well-defined stellar disc, this has the effect of
aligning the disc with the $X-Y$ plane, and the $Z$ coordinate
indicating height above the disc plane. In this coordinate system,
stars with height $|Z|<10$ kpc and cylindrical radius $R < 30$ kpc,
are selected for analysis. We post-process the snapshots to record the
positions of star particles, relative to the host galaxy center at
that time, in the first snapshot in which they appear. Since the
average time between snapshots is $\sim25 \Myr$ we can refer to this
quantity as the ``formation distance'' of the star particle without
much loss of fidelity. In summary, the analysis in this work uses the
disc-aligned coordinates, IMF-averaged metallicities, ages, and
formation distances of the selected stars.

As discussed in \citet{wetzel+16}, \citet{Hopkins2018fire2},
\citet{garrison-kimmel+18}, \citet{sanderson+18} and
\citet{elbadry+18a}, \thesim\ and the other MW analogs simulated in
this mass range with FIRE-2, have stellar-to-halo mass ratios and disc
properties resembling those of the MW and M31. In particular
\thesim\ has a thin gas disc and a double-exponential stellar disc
with comparable scale heights to the MW at $z=0$ (see Table
\ref{tbl:simdetails}). At the present day \thesim\ has about twice the
stellar mass of the MW.  It also has a much higher star formation
rate, even though about 50\% of its total stellar mass is in a
dispersion-supported system with the rest in a rotationally-supported
disc \citep{garrison-kimmel+18}, which is a significantly higher
dispersion-supported fraction than in the MW
\citep{bland-hawthorn_gerhard16}. The structure of the disc, bar, and
bulge in \thesim\ are emergent properties of the simulation, not the
result of tuned initial conditions. Thus we can confirm that triaxial
structures in the inner regions of galaxies can arise in a fully
cosmological formation scenario including filamentary accretion, the
response of a cold dark matter halo, and stellar feedback, though we
caution that AGN feedback is not included in this simulation.  This is
in good agreement with previous results from cosmological simulations
\citep{romano-diaz+08, scannapieco_athanassoula12, kraljic+12, goz+15,
fiacconi+15, okamoto+15, bonoli+16, spinoso+17, buck+18}.

\begin{table*}
\caption{Structural properties of \thesim.}
\begin{center}
\begin{tabular}{lrl}
\hline
Property & Value & Unit\\
\hline
\hline
$M_h$ (halo mass at $z=0$; \citet{bryan_norman98}) & $1.6 \times 10^{12}$ & $\Msun$\\
$M_*$ (stellar mass at $z=0$)	& $1.1 \times 10^{11}$ &  $\Msun$\\
$M_{\mathrm{gas}}$ (gas mass at $z=0$) &  $1.4\times 10^{10}$ &  $\Msun$\\
\hline
Baryon particle mass	& 7070 & $\Msun$\\
Dark matter particle mass	& $3.52 \times 10^4$ & $\Msun$\\
Dark matter softening length & 40 & \pc \\
Star softening length & 4.0 & \pc \\
Gas smoothing / softening (minimum) & 1.0 & \pc\\
\hline
$R^*_{90}$ (2D radius enclosing 90\% of $M_*$) & 13.3 & \kpc\\
$Z^*_{90}$ (height enclosing 90\% of $M_*$) & 2.75 & \kpc\\
$R_{\mathrm{gas}}$ (defined in \citealt{garrison-kimmel+18}) & 12.1 & \kpc\\
$Z_{\mathrm{gas}}$ (defined in \citealt{garrison-kimmel+18}) & 656 & \pc\\
\hline
scale height of thin stellar disc at $8.2 \pm 0.2 \kpc$ & 380 & \pc\\
scale height of thick stellar disc at $8.2 \pm 0.2 \kpc$ & 1240 & \pc \\
scale height of cold ($T < 100$K) gas disc & 260 & \pc \\
star formation rate at $z=0$ & 7.5 & $\Msun$/yr \\
\hline
\hline
\end{tabular}
\end{center}
\label{tbl:simdetails}
\end{table*}

\subsection{Simulation scaling}
\label{ss:rescaling}

\thesim\ has a bar of semi-major axis $a_B \simeq 6\kpc$. Through 
most of the paper we present \thesim\ without rescaling it.  However,
studying the vertex deviation and the age distribution requires
\thesim\ to be scaled in size such that the vertical structure is
comparable to that in the MW.
We do this by placing the arms of the X-shape in \thesim\ at $z=0$, as
traced by the peaks in the line-of-sight density distribution, at a
comparable location as in the MW.  We compute the factor required to
obtain a half-length of $\sim$ 2 kpc for the X-shaped bulge in
\thesim.  We find that a factor of 0.5, applied to all particles,
accomplishes this and results in the arms of the X-shaped bulge having
a similar size to those of the MW bulge as mapped by
\citet{wegg_gerhard13}.  To further ensure that this scaling is
suitable for comparing \thesim\ to the MW, we measure the distance
distribution of all stars along the minor axis at different latitudes
to identify the Galactic latitude at which the split in distance
distributions is first identified. We find that when using a scaling
factor of 0.5 the split is first seen at a latitude of $\rm
b\sim5\degrees$, which compares well with the MW's bulge
\citep[c.f.][]{mcwilliam_zoccali10}.  The Sun is then placed at $8
\kpc$ from the Galactic centre and the bar is rotated to an angle of
$27\degrees$ with respect to the Galactic centre-Sun line of sight.

\begin{figure}
\includegraphics[width=0.45\textwidth,angle=0,trim=0.0cm 0.0cm 0.8cm 0.0cm,clip=true]{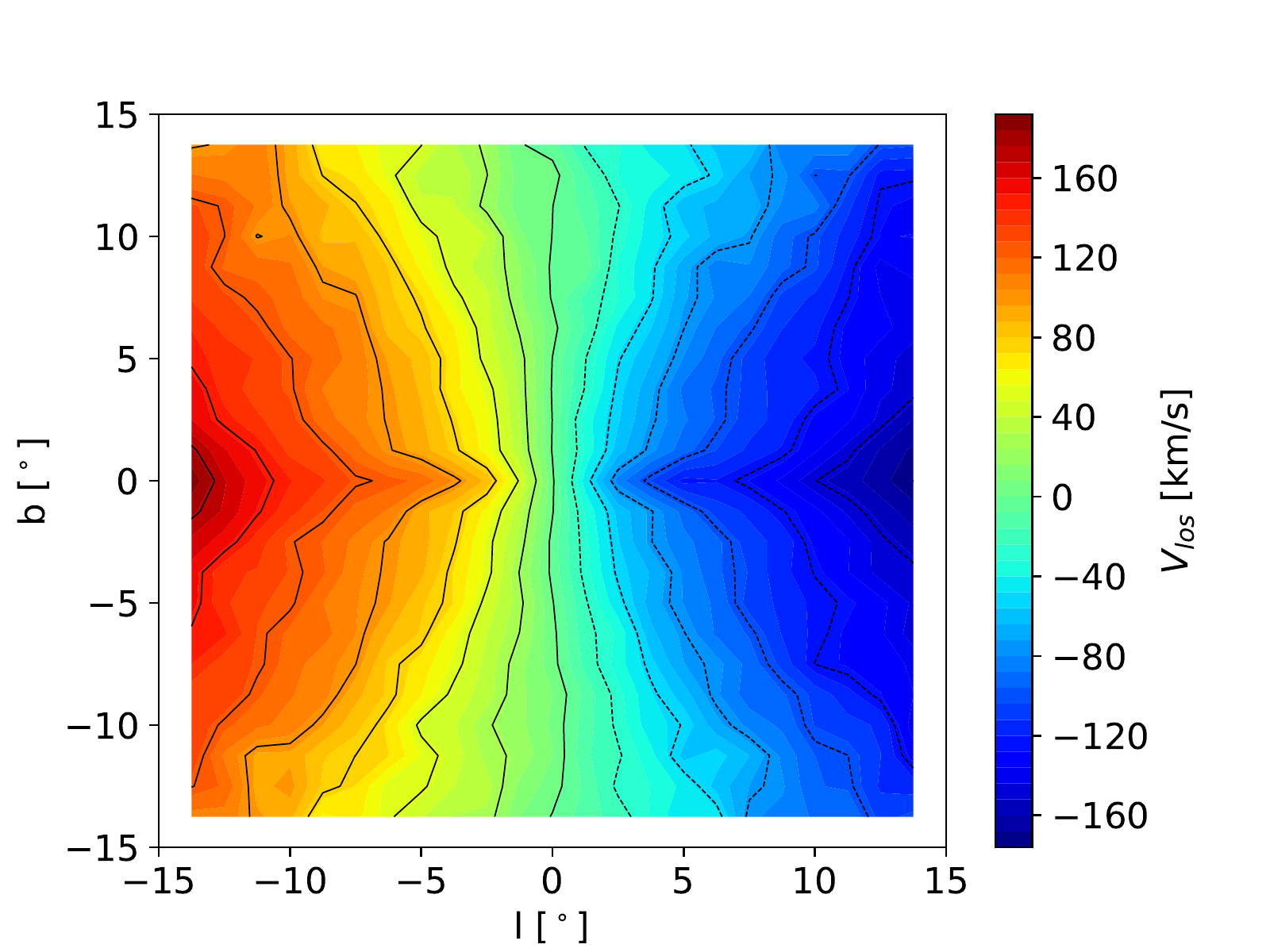}\\
\includegraphics[width=0.45\textwidth,angle=0,trim=0.0cm 0.0cm 0.8cm 0.0cm, clip=true]{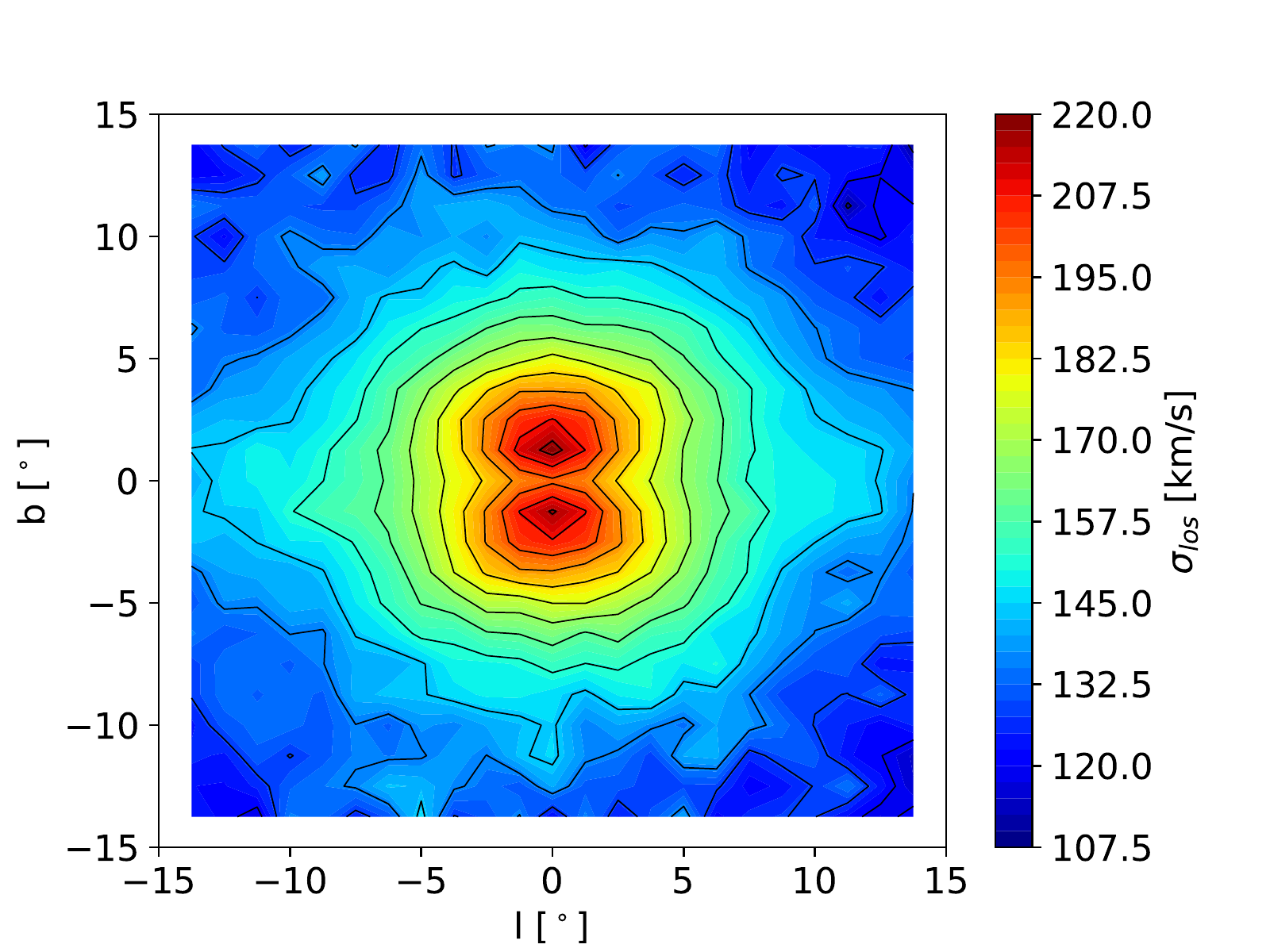}
\caption{Mean line-of-sight velocity (top) and velocity dispersion
  (bottom) maps of the bulge region for stars at a distance from the
  Sun $6 < R/\kpc < 10$ in \thesim\ when rescaled to the MW as
  described in Section \ref{ss:rescaling}.}
\label{f:2d_vel_maps}
\end{figure}

We apply no scaling to the velocities because none are needed for our
analysis; for the vertex deviation analysis, we are only interested in
ratios of dispersions, which do not require the model to be
kinematically scaled to the MW.  We present maps of the mean velocity
and velocity dispersion along the line of sight in Galactic
coordinates in Fig. \ref{f:2d_vel_maps}. The kinematic maps show
a (close-to) cylindrical rotation and a clear peak in velocity
dispersion in the central regions that appears vertically
elongated. These properties are in good qualitative (but not
quantitative) agreement with the ones observed in the MW
\citep{zoccali+14, ness+16b}, in simulations \citep{qin+15,
fragkoudi+17, buck+18}, and in similar external galaxies
\citep{gonzalez+16, molaeinezhad+16}. In particular, the vertically
elongated velocity dispersion `peak' presented in \citet{zoccali+14}
is clearly observed in the simulation once it is rescaled.

\subsection{Star formation history and chemistry}
\label{ss:mdfsfh}

\begin{figure}
\centerline{
\includegraphics[angle=0.,width=\hsize]{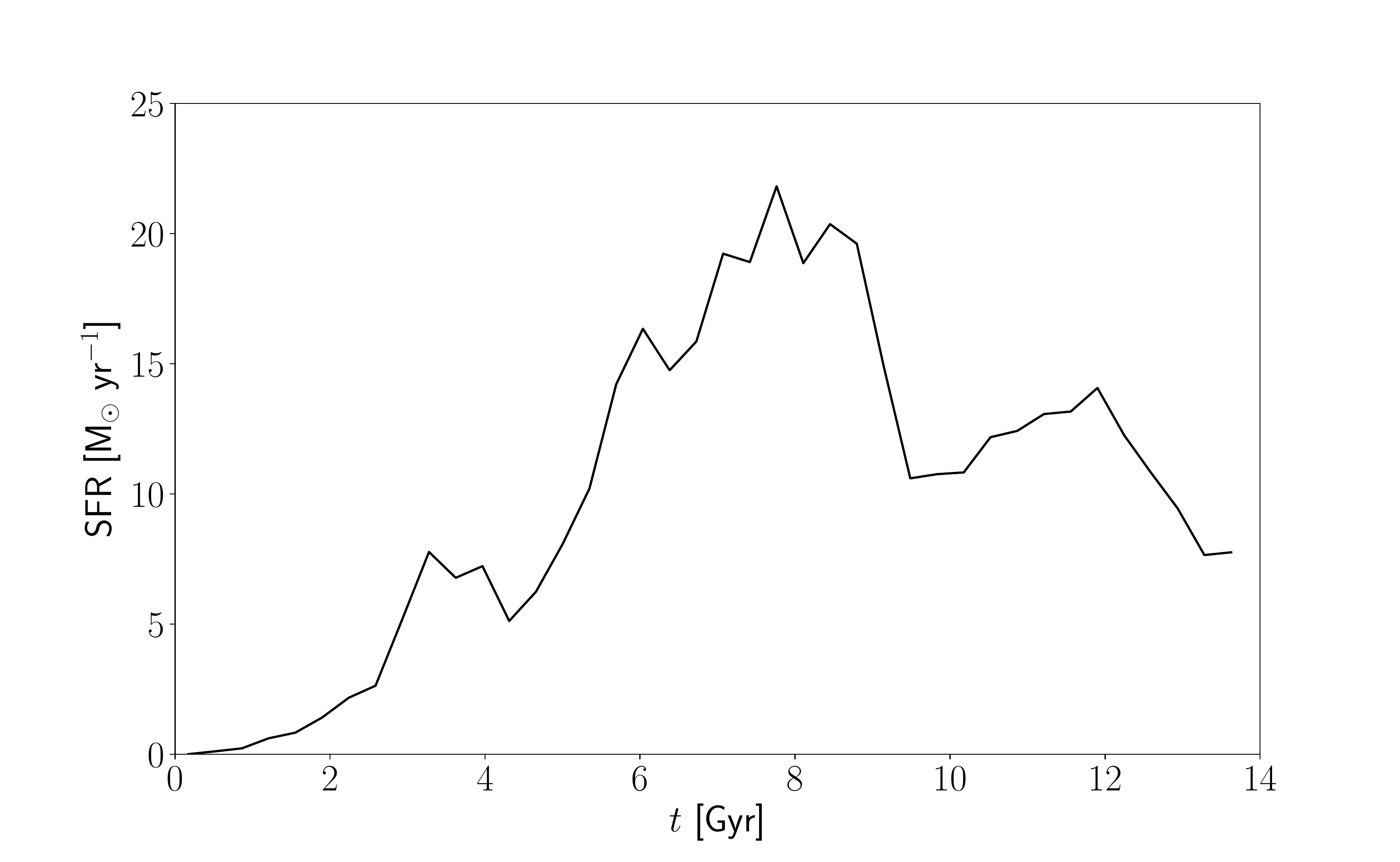}
}
\centerline{
\includegraphics[angle=0.,width=\hsize]{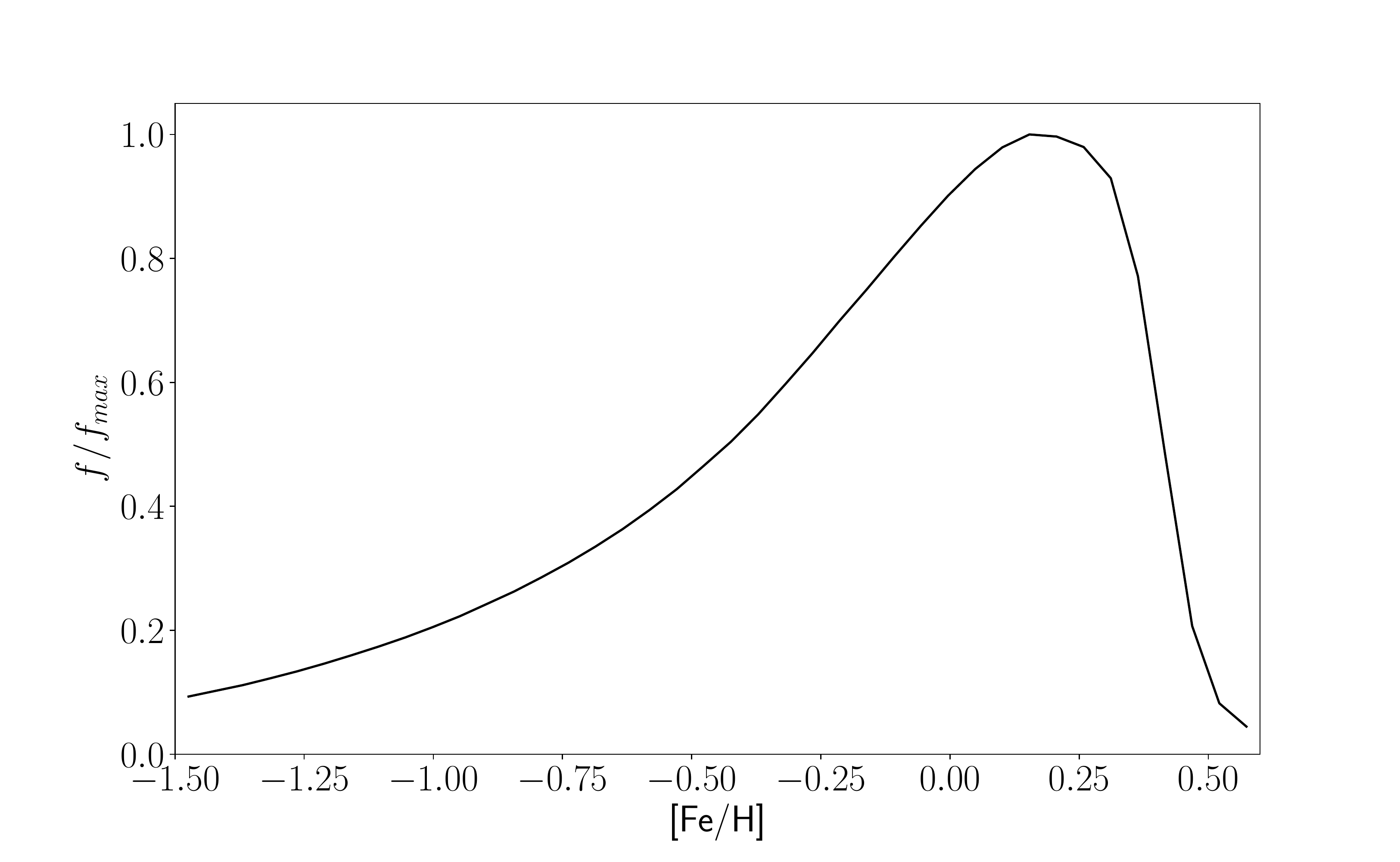}
}
\caption{Top: The star formation history of the model. The initial 
star formation rate is very low but it rises rapidly to a peak at
$\sim 8\Gyr$.  Bottom: The metallicity distribution function over the
entire galaxy.
\label{fig:mdfsfh}}
\end{figure}

The top panel of Fig. \ref{fig:mdfsfh} shows the star formation
history of the model.  Star formation is initially very low, peaking
at around $8\Gyr$.  The star formation rate of the accreted component
(not shown) peaks at $\sim 3\Gyr$, at which time it accounts for $\sim
40\%$ of the total star formation rate; essentially no star formation
occurs after $4 \Gyr$ in the accreted component.  By $t\sim 9\Gyr$ the
star formation rate has dropped by roughly a factor of two.  A second
drop, again by a factor of two, in the star formation rate occurs
shortly before $t = 12\Gyr$.

The bottom panel of Fig. \ref{fig:mdfsfh} shows the metallicity
distribution function (MDF) across the model.  The model's MDF peaks
at nearly Solar metallicity and has a long tail to low \feh.  The MDF
is similar to some extent to the MDF of the bulge
\citep[e.g.][]{zoccali+08, gonzalez+15b}.


\section{Kinematic Fractionation}
\label{s:kfractionation}

\begin{figure}
\centerline{
\includegraphics[angle=0.,width=\hsize]{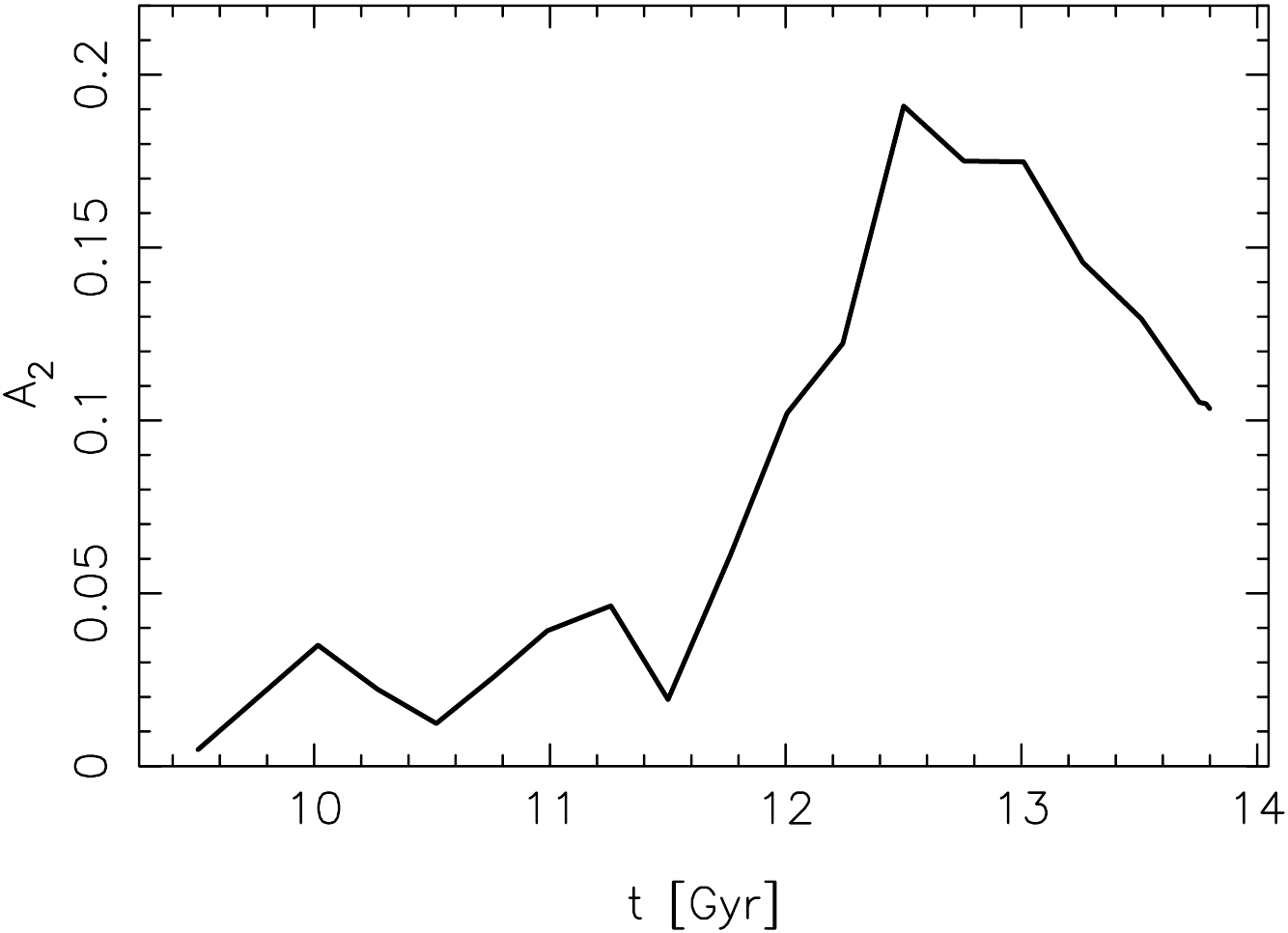}
}
\caption{Bar amplitude evolution in \thesim\ since redshift $z = 0.4$
  (corresponding to $4.8 \Gyr$ of evolution).
  \label{fig:barformation}}
\end{figure}

Fig. \ref{fig:barformation} shows the evolution over the last $4.8
\Gyr$ (\ie\ since redshift $z = 0.4$), of the bar amplitude, $A_2$,
defined as the usual $m=2$ amplitude of the Fourier moment measured
over all stars \citep[e.g.][]{debattista_sellwood00}.  The bar forms
quite late, starting from $11.5\Gyr$ ($z\simeq 0.19$).  It reaches a
peak amplitude at $\sim 12.7\Gyr$, and weakens somewhat in the next
\Gyr, as is often seen in simulations of isolated galaxies.  In
isolated simulations, bars generally experience renewed growth past
this point \citep[e.g.][]{combes_sanders81, sellwood_moore99,
  debattista_sellwood00, bournaud_combes02, oneill_dubinski03,
  athanassoula02, martinez-valpuesta+06}, but the late bar formation
in this simulation does not give the bar time to strengthen again.

\subsection{Density separation}
\label{ss:density}

\begin{figure*}
\centerline{
\includegraphics[angle=0.,width=0.5\hsize]{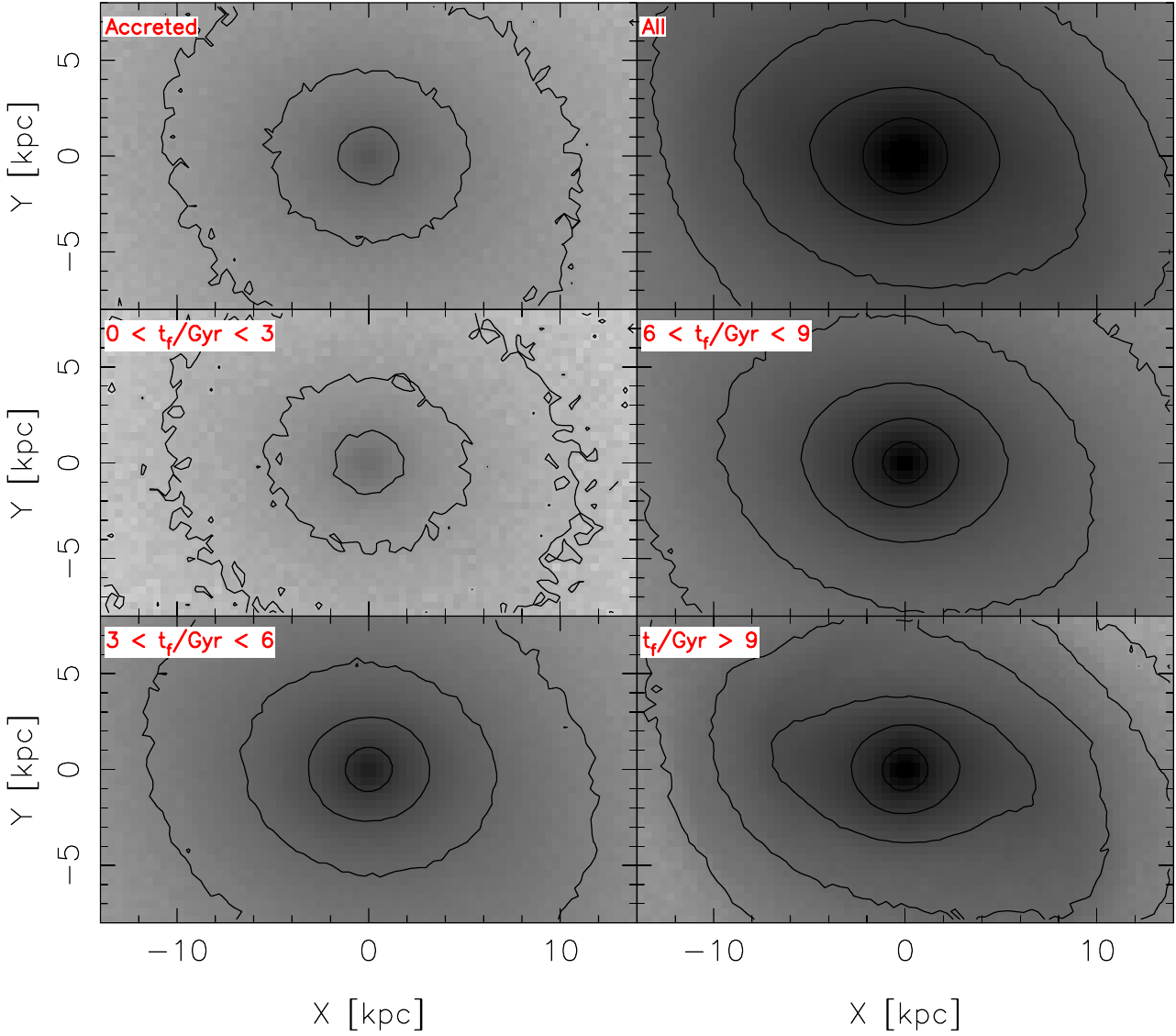}
}
\centerline{
\includegraphics[angle=0.,width=0.5\hsize]{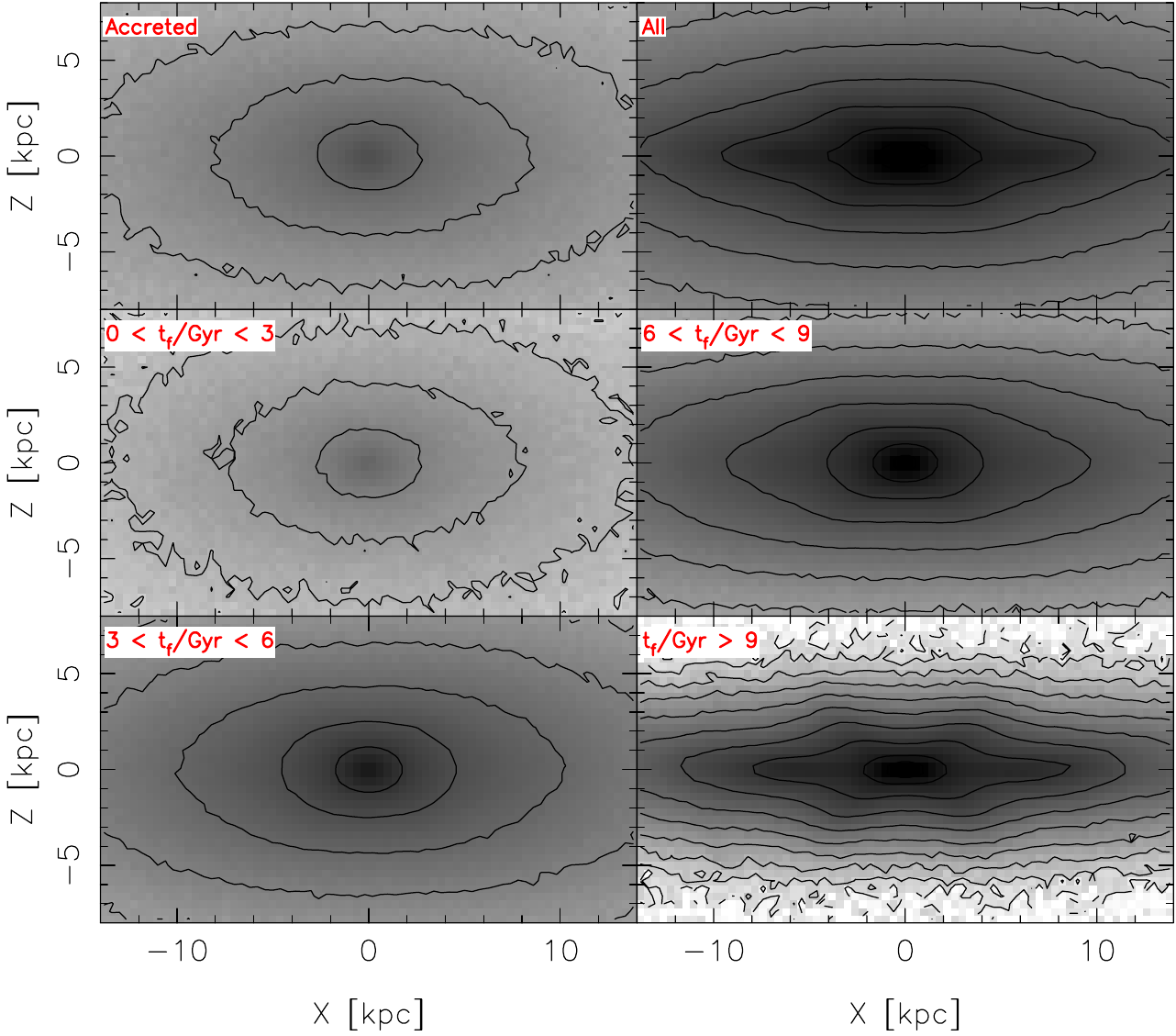} 
\includegraphics[angle=0.,width=0.5\hsize]{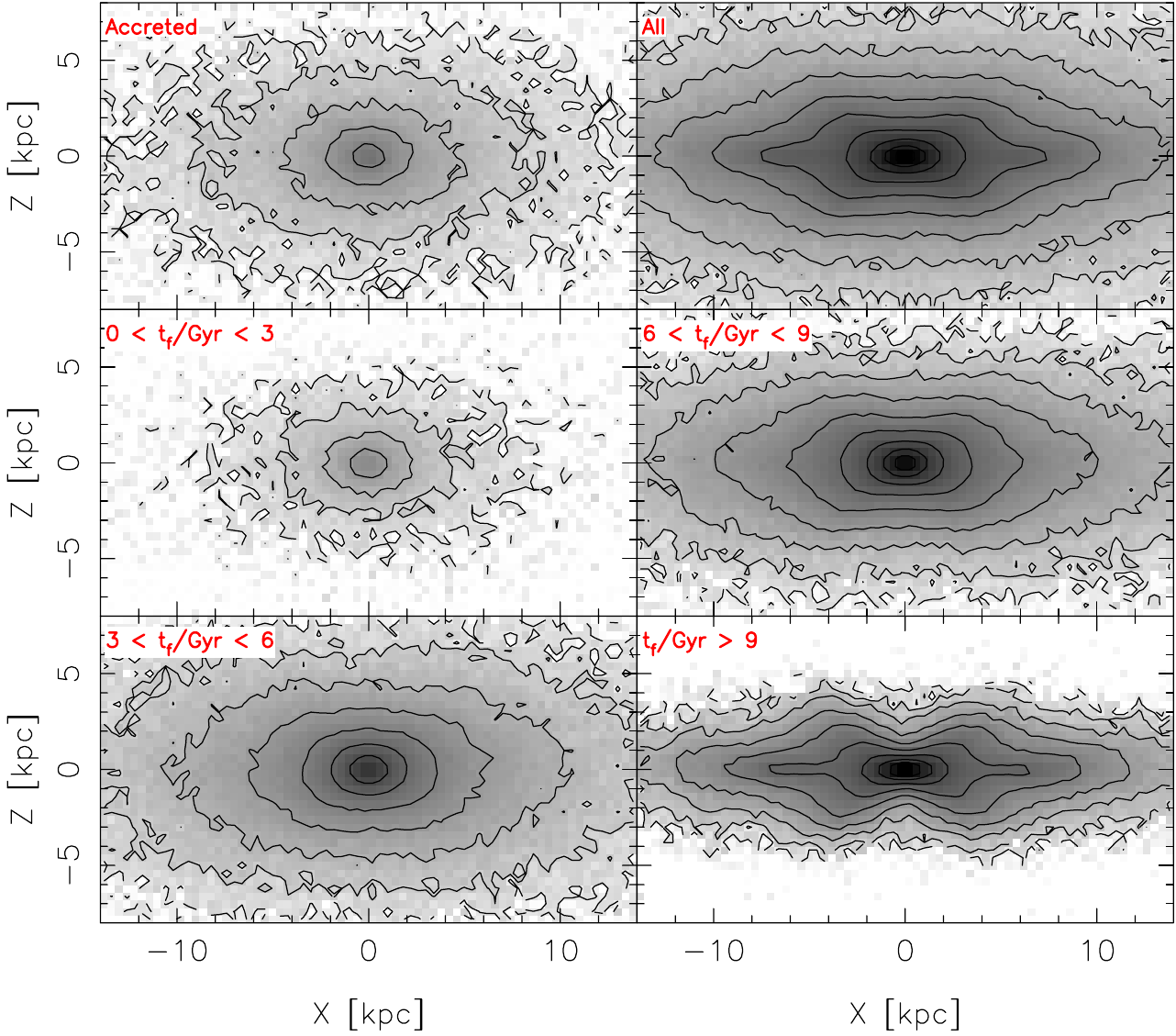} 
}
\caption{Density distributions in the face-on (top panels) and edge-on
  (bottom panels) projections.  The bottom left panels show the full
  edge-on projection while the bottom right panels show a
  cross-section with particles chosen within $|Y| \leq 0.25 \kpc$.  In
  the top row of each set of panels we show the accreted stars (left)
  and all the stars (right).  The rest of the panels show populations
  separated by time of formation of the stars, $\tform =13.8 \Gyr -$age.
  \label{fig:densitybyage}}
\end{figure*}

Fig. \ref{fig:densitybyage} shows the mass density distribution at $z
= 0$ in face-on and edge-on projections for the model separated by
different stellar populations.  As in \citet{debattista+17}, younger
populations exhibit a stronger bar, and a more prominent box/peanut
(B/P) shape, than the older ones.  The difference in the B/P strength
as a function of age is a signature of kinematic fractionation, as
discussed in \citet{debattista+17}.

Fig. \ref{fig:densitybyage} also shows the density distribution of the
stellar population that was accreted, which we define as stars that
formed at radius $r_\mathrm{f} > 40 \kpc$.  The accreted stars, which
account for $4.7\%$ of all the stars, formed primarily ($> 98\%$) by
redshift $z = 1.27$ ($\tform=5\Gyr$).  They have a density distribution
similar to that of the oldest (age $ > 10.8\Gyr$ now) in-situ stars,
\ie\ those formed at $r_\mathrm{f} < 40 \kpc$ \citep[see
  also][]{elbadry+18b}.  Like the oldest bin, no bar or X-shape is
present in the accreted population.

\subsection{Age and metallicity dependence of bar amplitude}
\label{ss:baramp}

\begin{figure}
\centerline{
\includegraphics[angle=0.,width=\hsize]{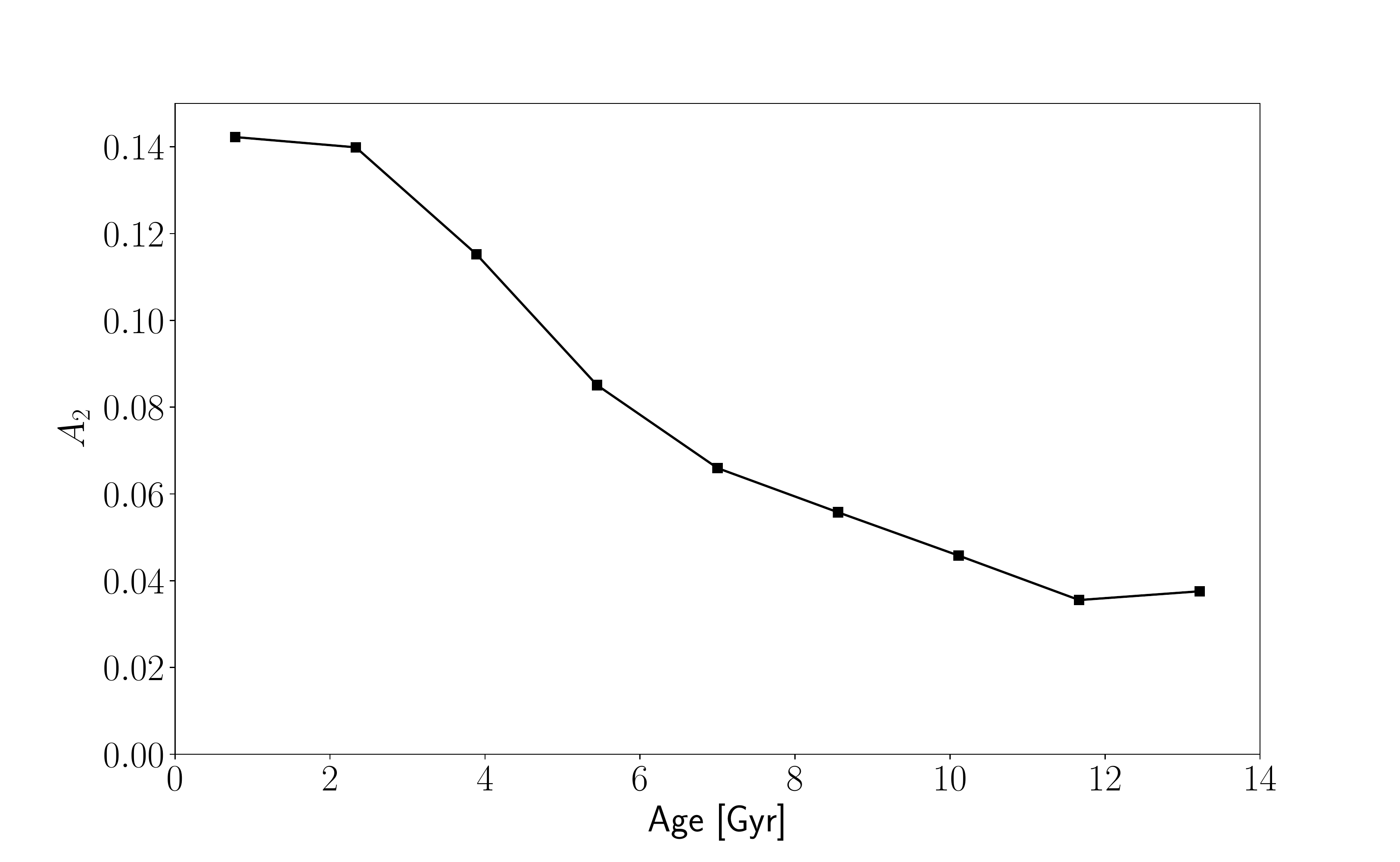}
}
\centerline{
\includegraphics[angle=0.,width=\hsize]{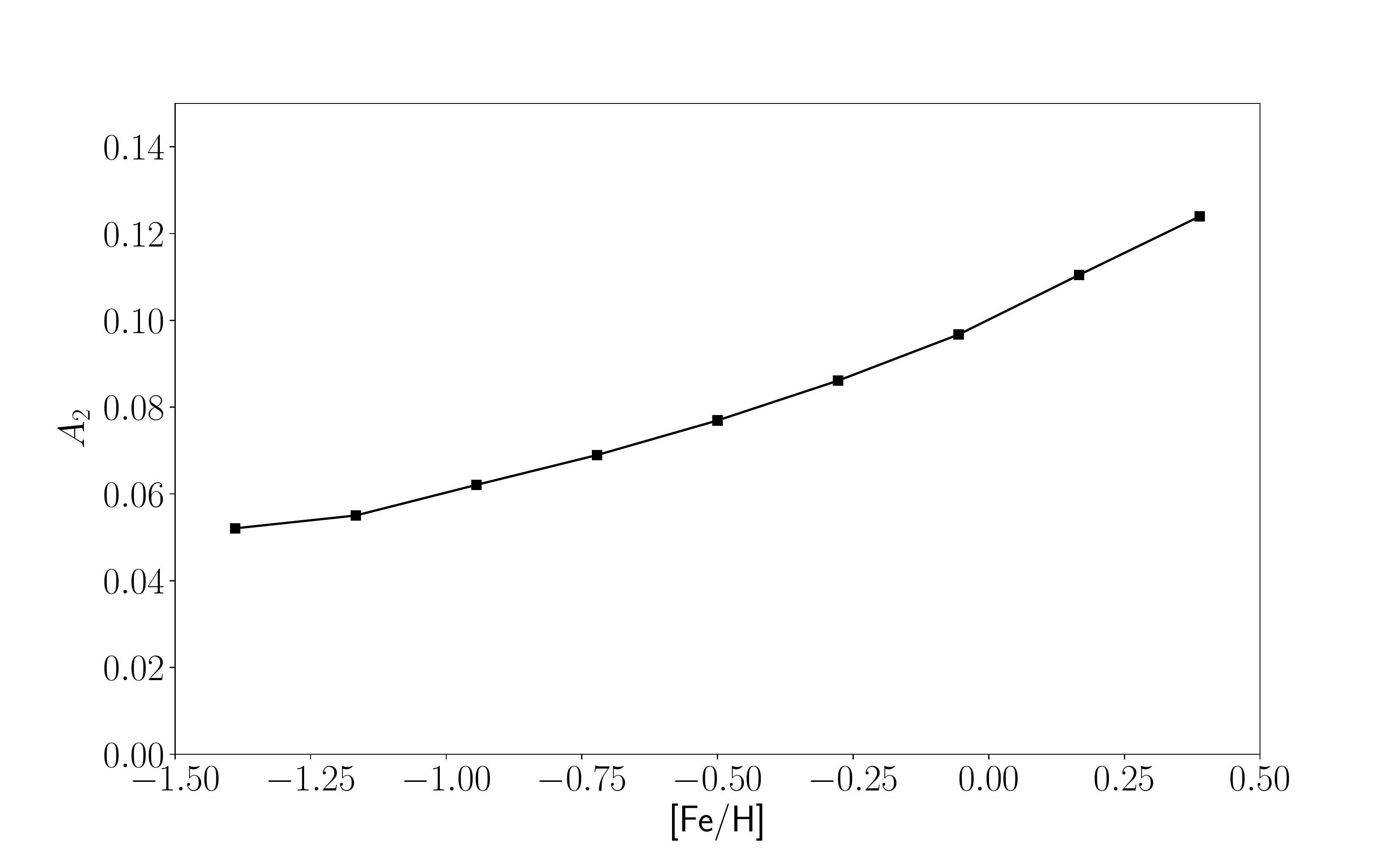}
}
\caption{Bar amplitude as a function of stellar age (top) and 
metallicity (bottom) at $z=0$.
\label{fig:baramps}}
\end{figure}

Fig. \ref{fig:baramps} shows the dependence of the usual $m=2$ Fourier
global bar amplitude on the age and metallicity.  As also apparent
from the face-on maps in Fig. \ref{fig:densitybyage}, the bar is
stronger in younger populations than in the older ones.  The very
oldest populations have a quite weak bar overall, while the youngest
populations have a $3.5\times$ stronger bar.

A comparable trend can be seen in the dependence of the bar amplitude
on the metallicity.  The weakest bar is found in the most metal-poor
stars while the most metal-rich stars have the strongest bar.  The
range of bar amplitudes spanned by the metallicity range is comparable
to that spanned by ages, and is continuously varying.  

These two trends in the behaviour of bar amplitude are consistent with
the results of \citet{debattista+17} and \citet{buck+18}.

\subsection{Deconstructing the X-shape by age}
\label{ss:xage}

\begin{figure*}
\centerline{
\includegraphics[angle=0.,width=0.33\hsize]{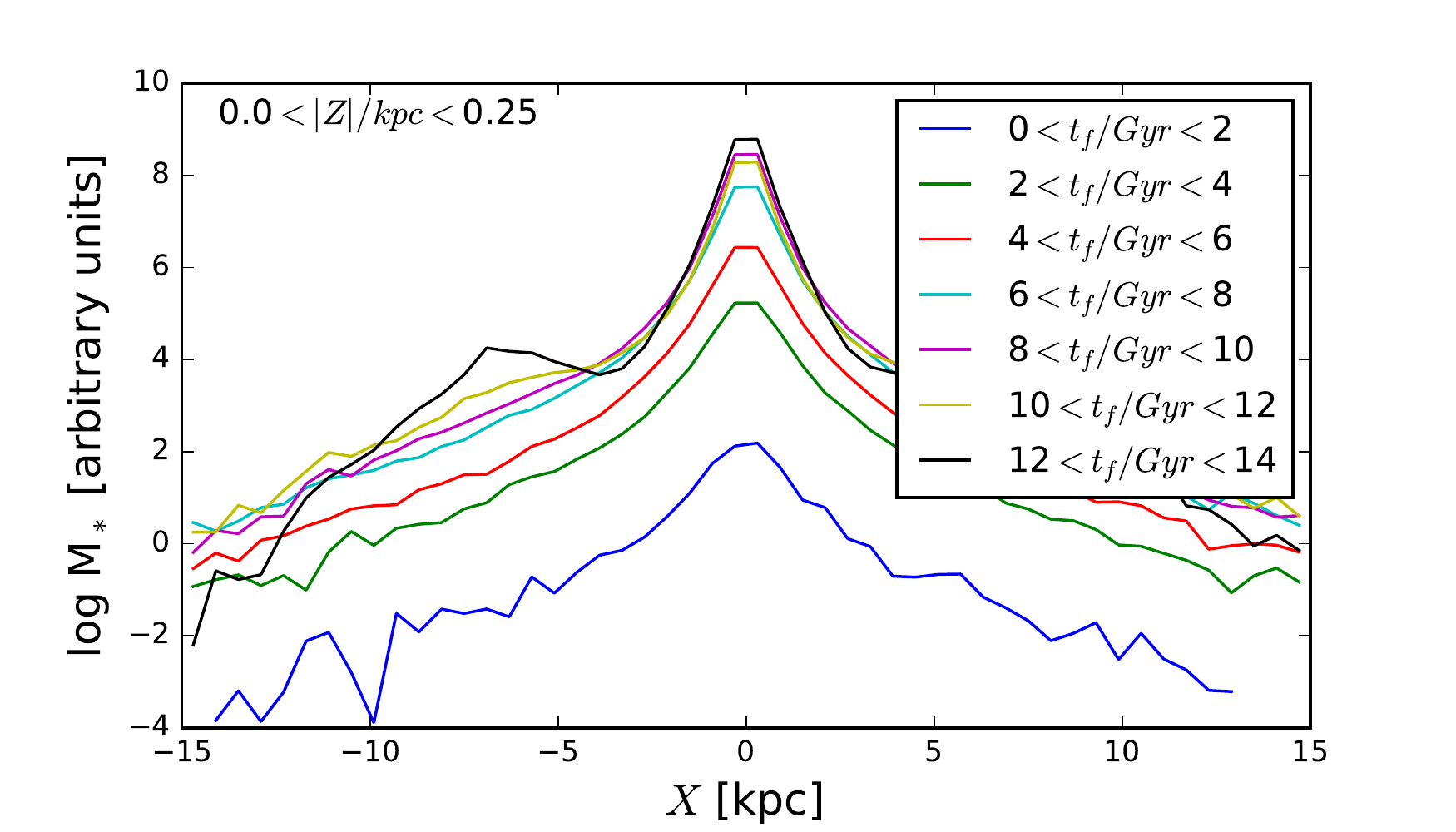}
\includegraphics[angle=0.,width=0.33\hsize]{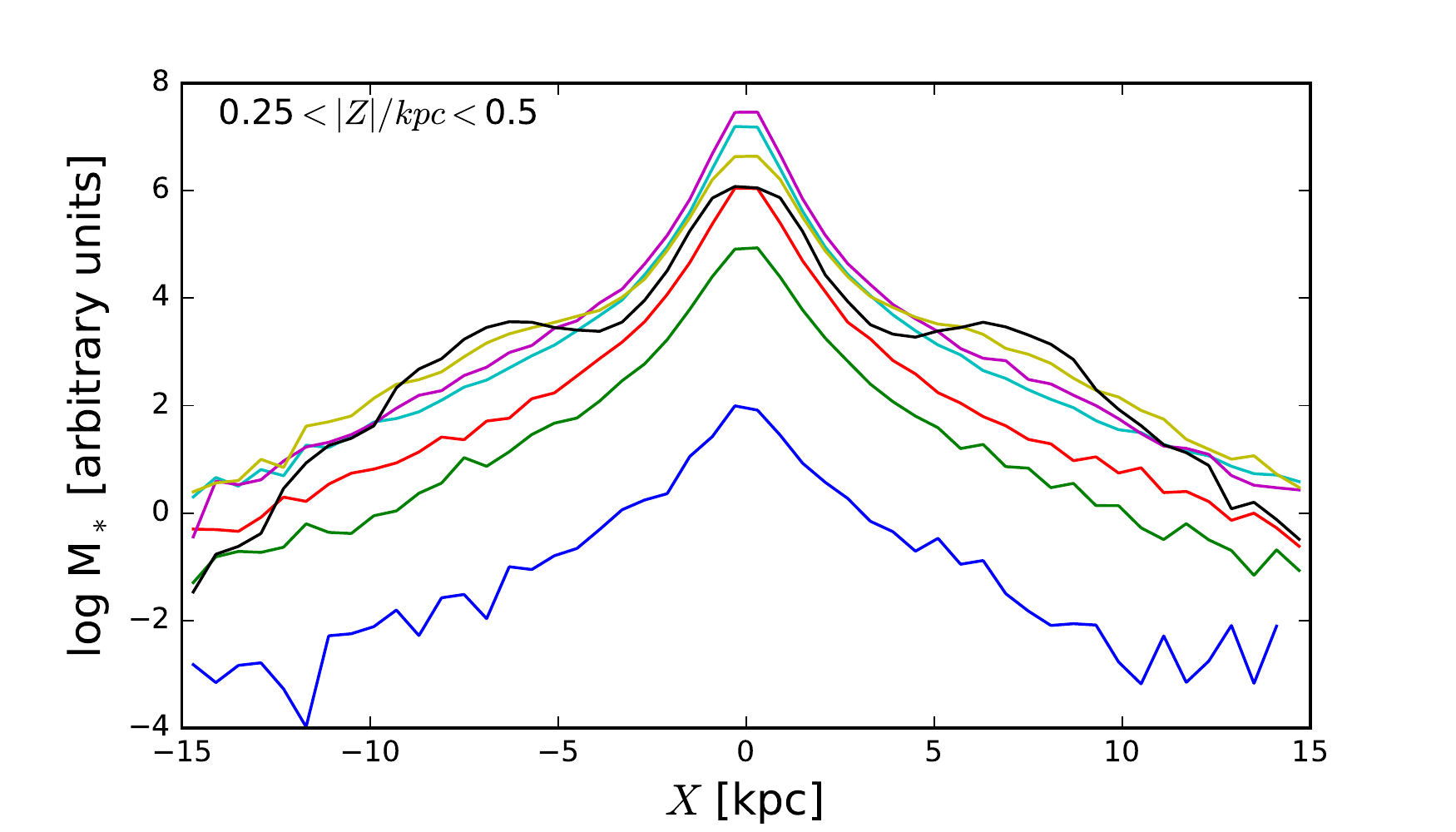}
\includegraphics[angle=0.,width=0.33\hsize]{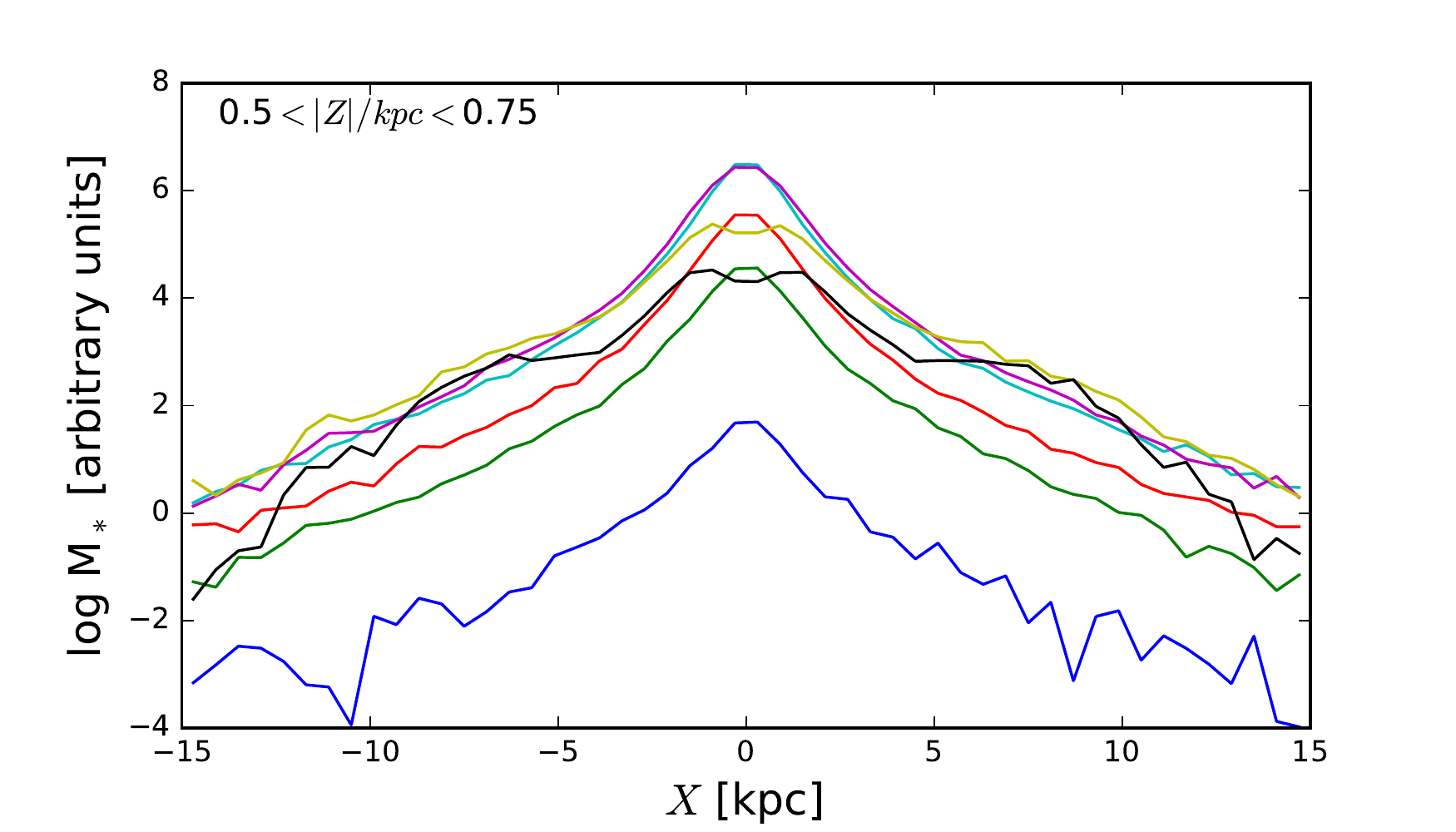}
}
\centerline{
\includegraphics[angle=0.,width=0.33\hsize]{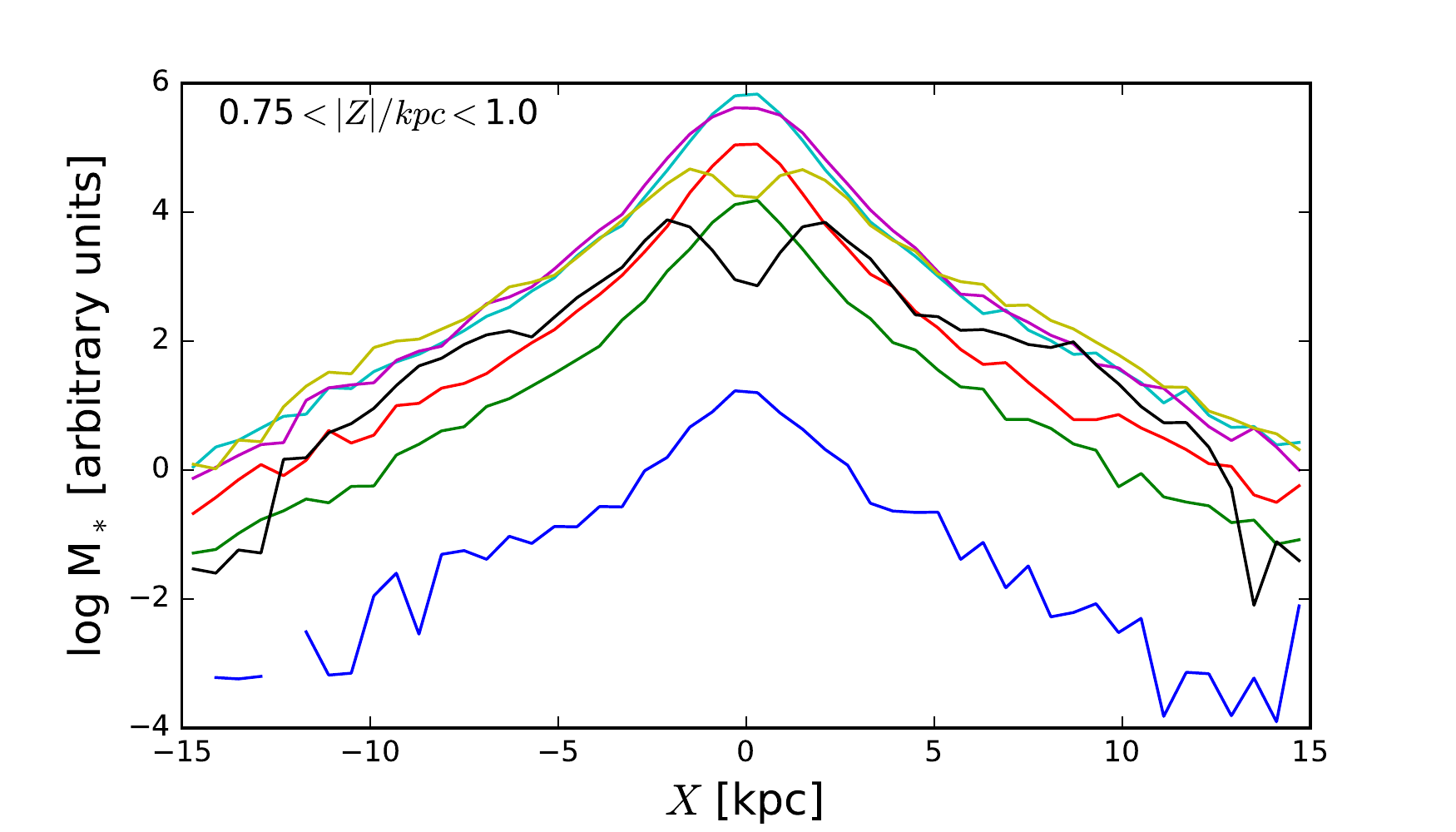}
\includegraphics[angle=0.,width=0.33\hsize]{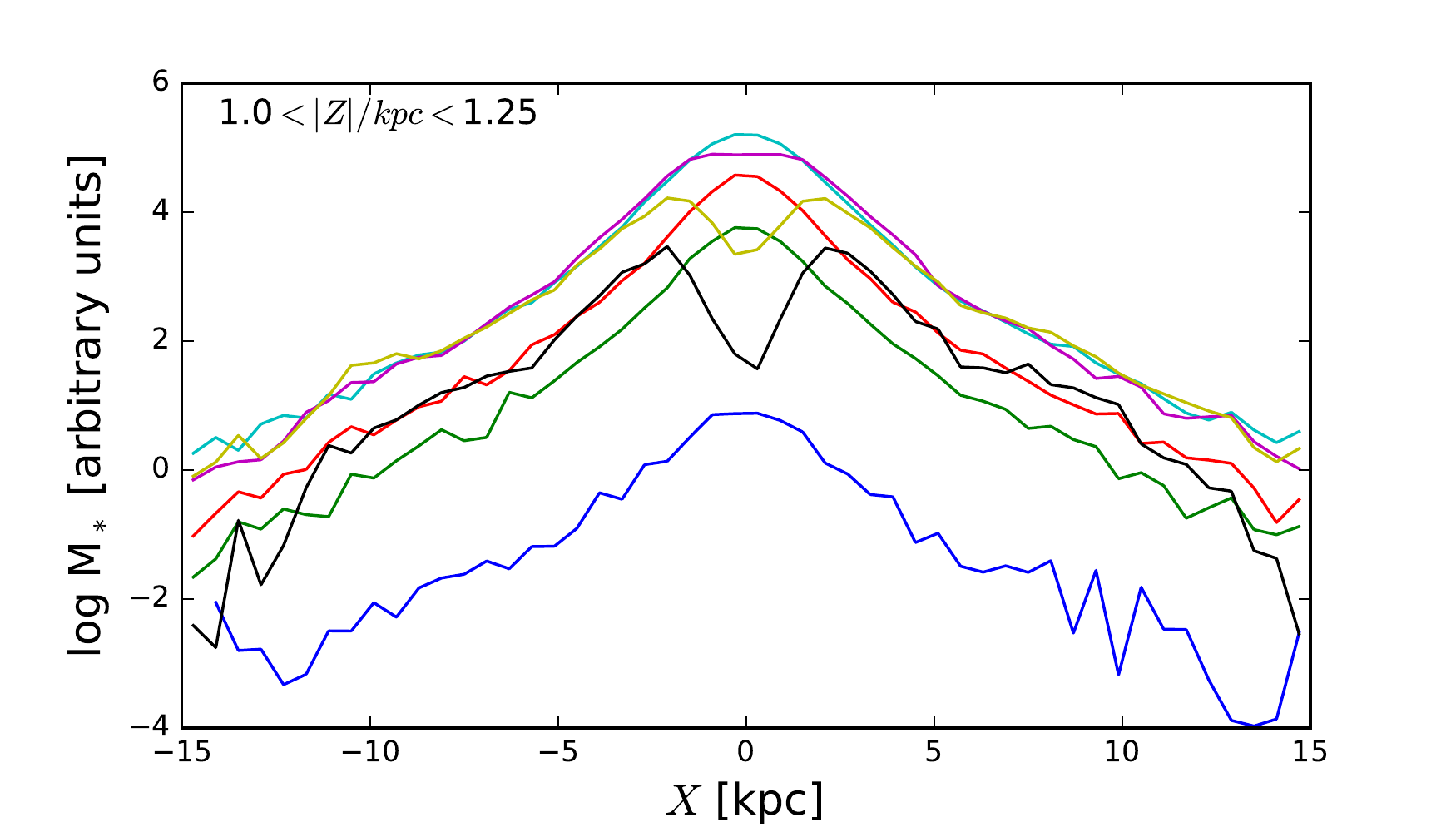}
\includegraphics[angle=0.,width=0.33\hsize]{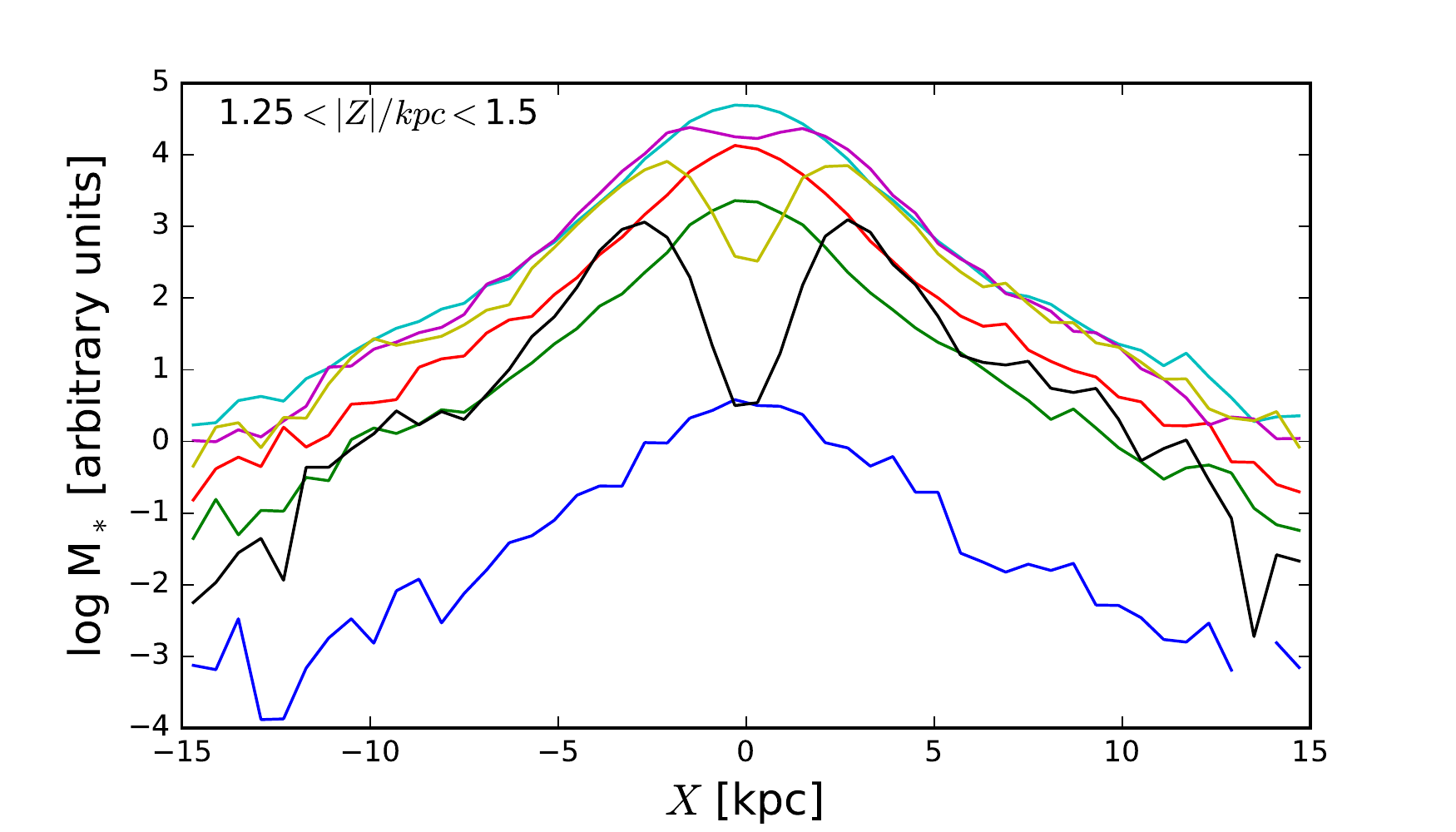}
}
\centerline{
\includegraphics[angle=0.,width=0.33\hsize]{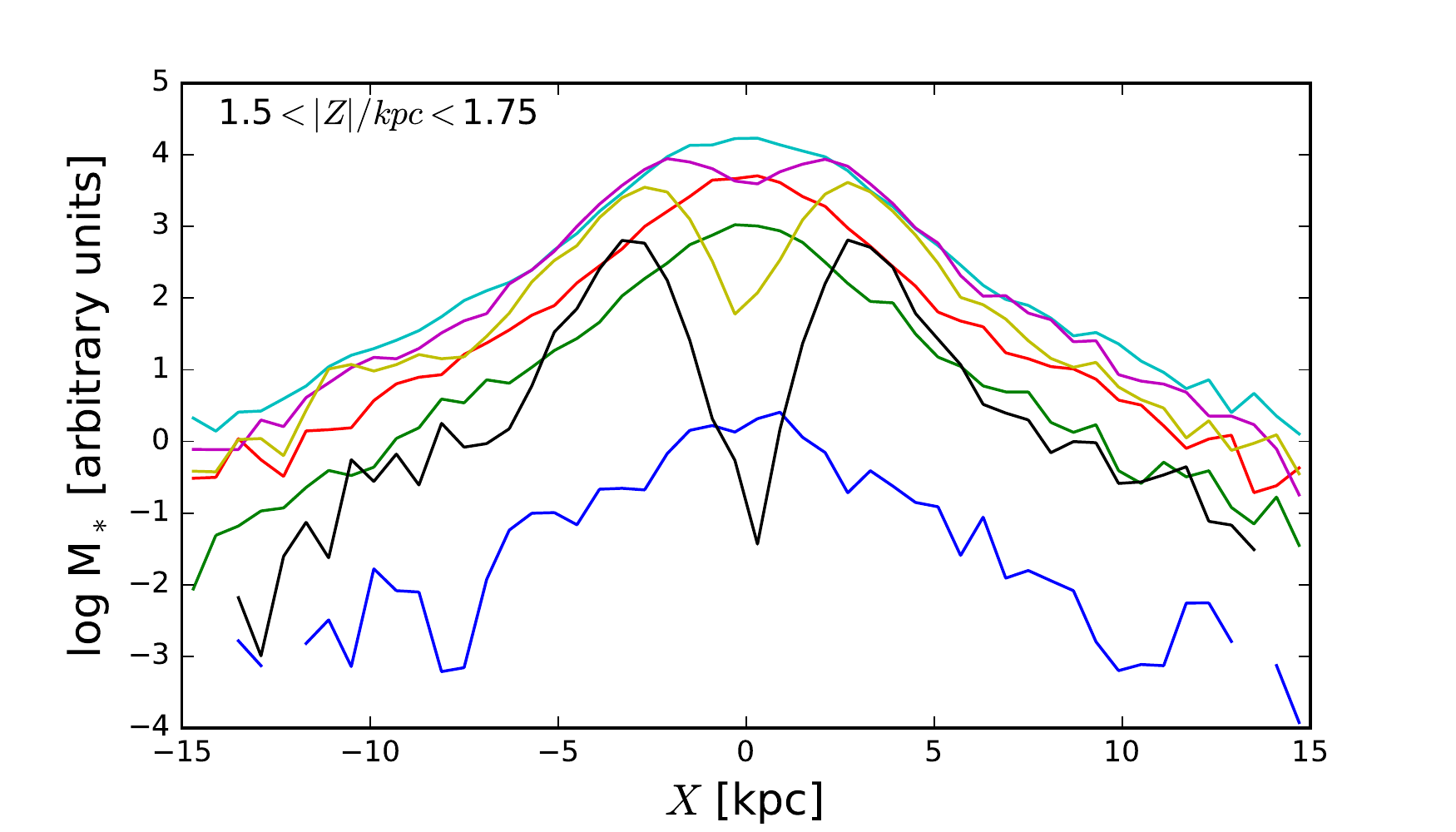}
\includegraphics[angle=0.,width=0.33\hsize]{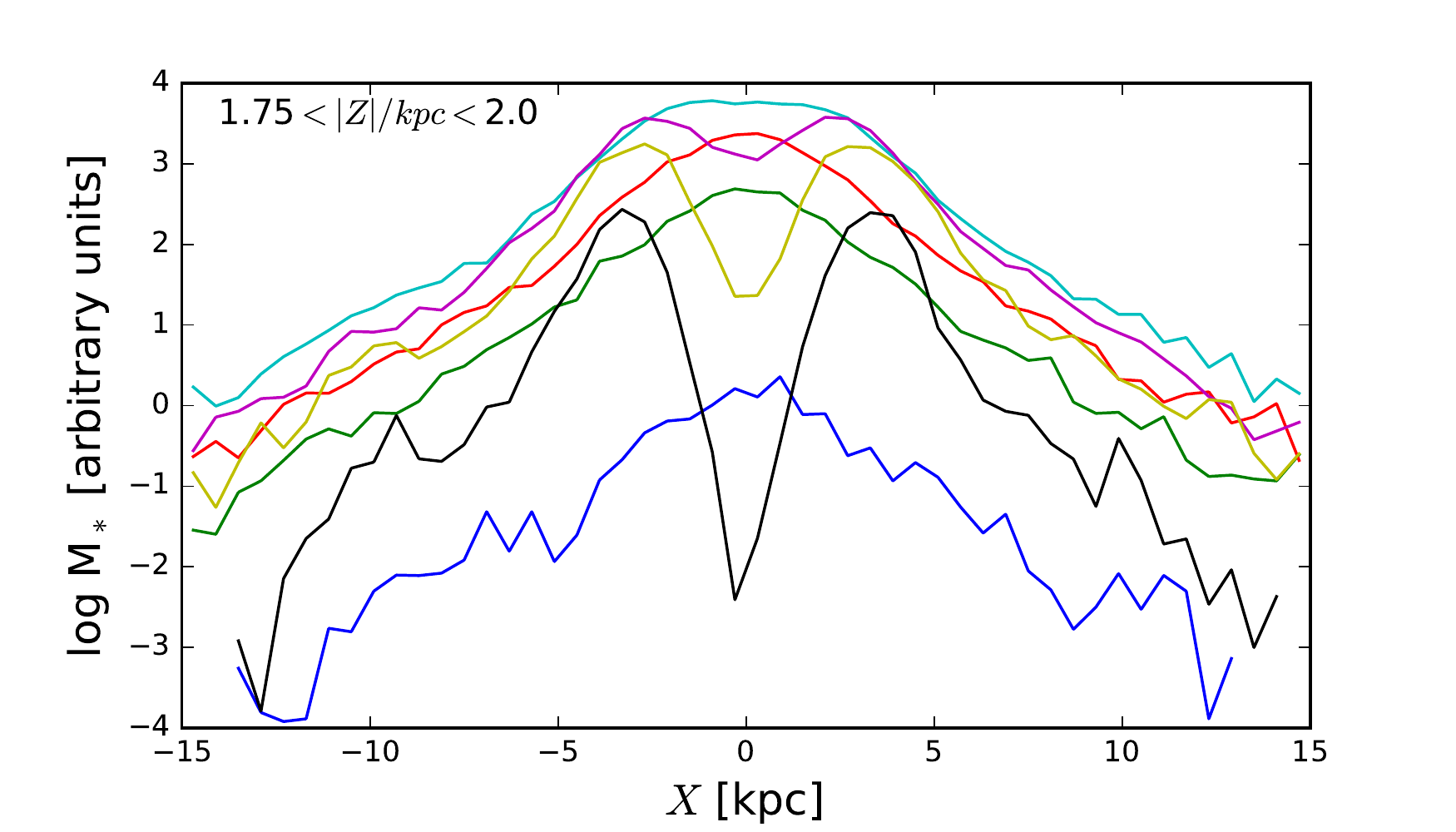}
\includegraphics[angle=0.,width=0.33\hsize]{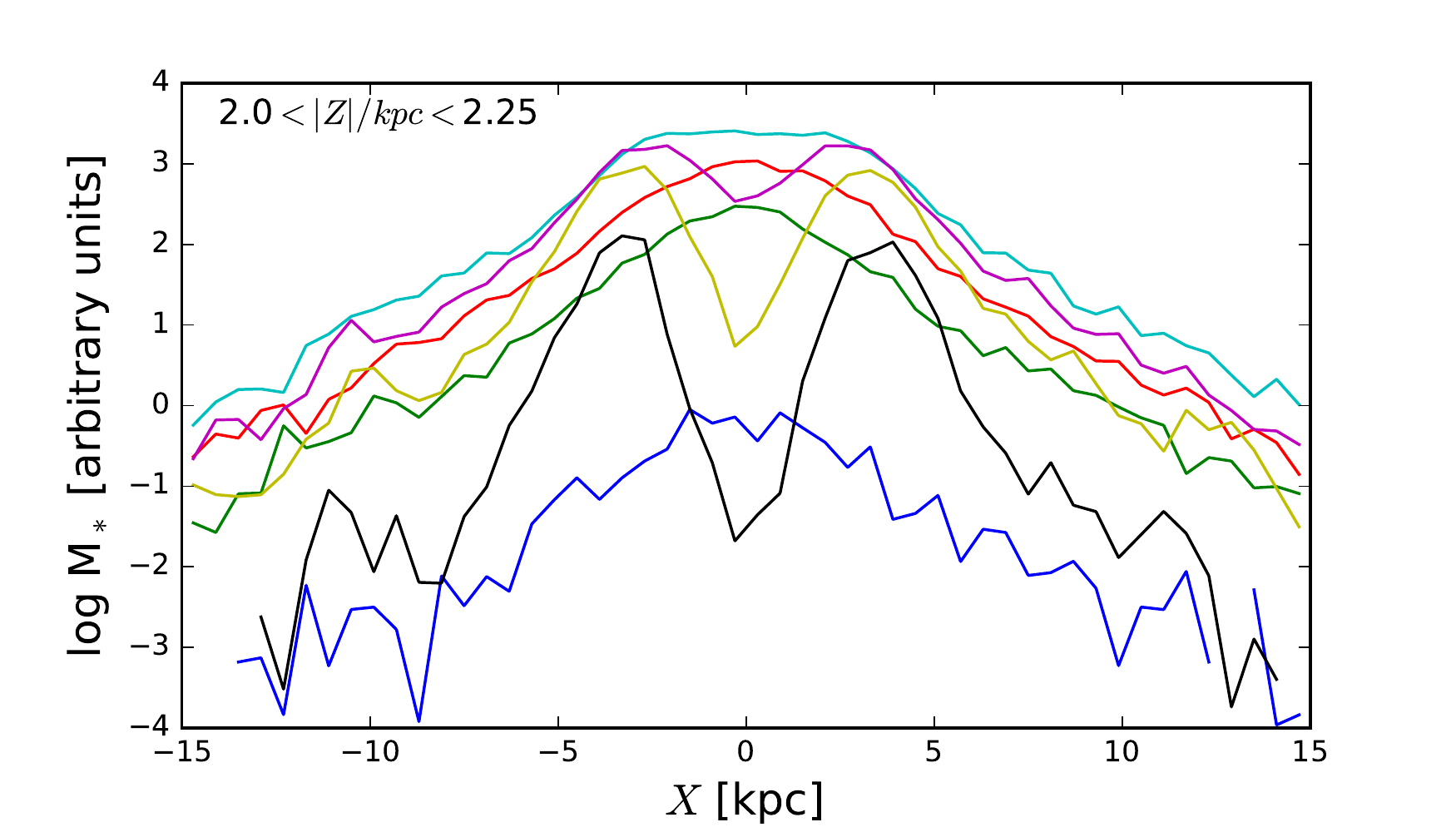}
}
\centerline{
\includegraphics[angle=0.,width=0.33\hsize]{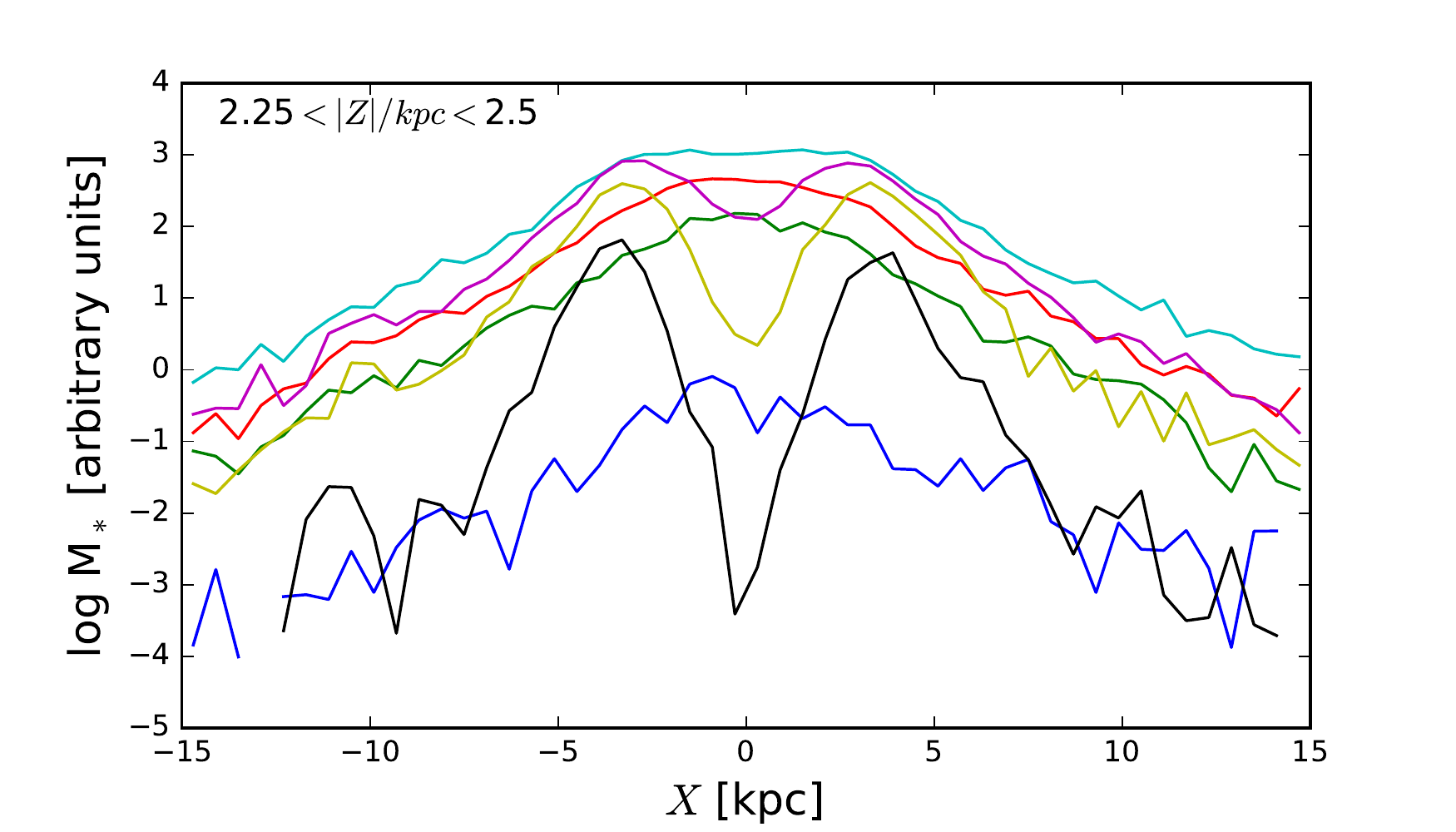}
\includegraphics[angle=0.,width=0.33\hsize]{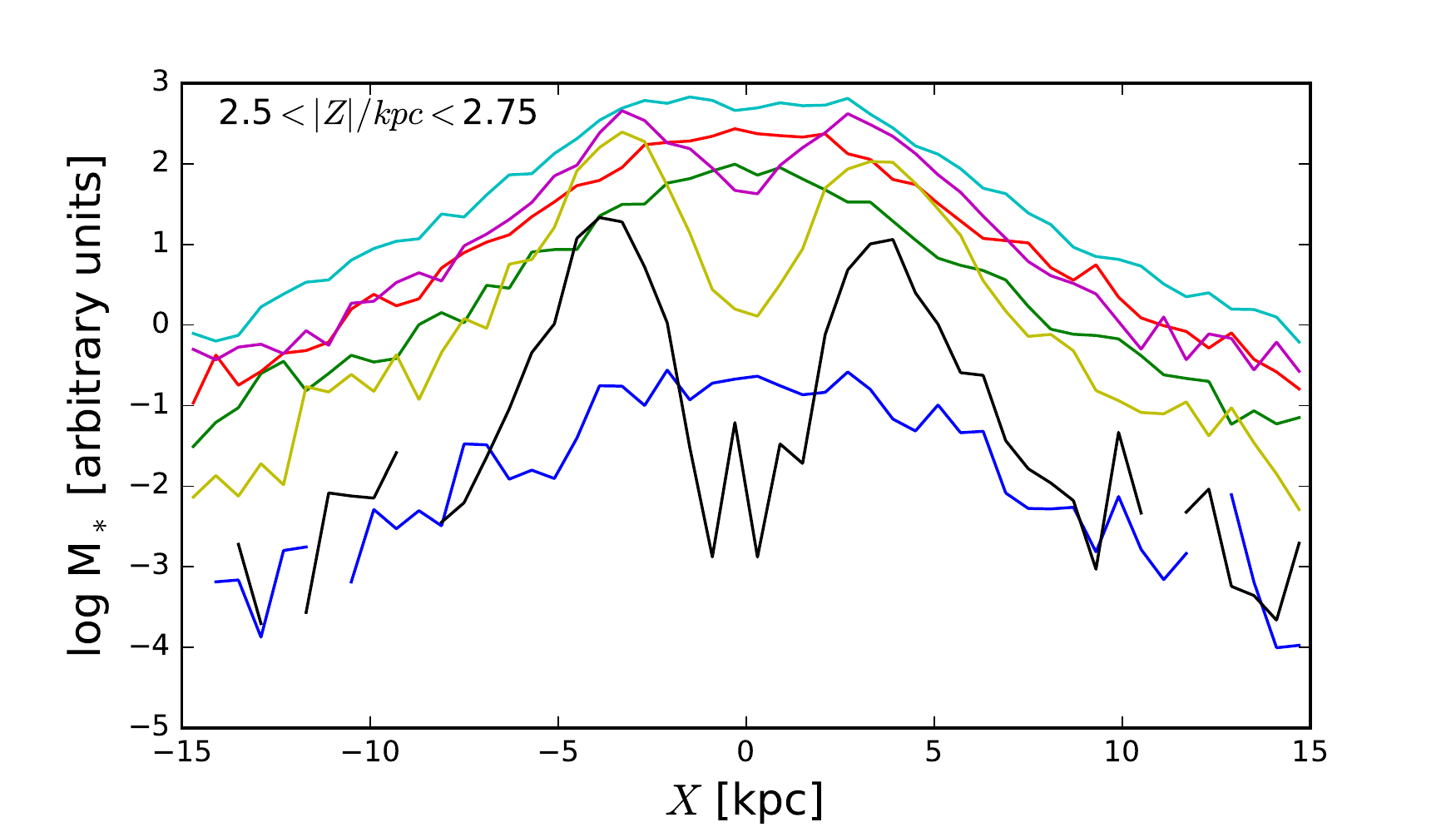}
\includegraphics[angle=0.,width=0.3\hsize]{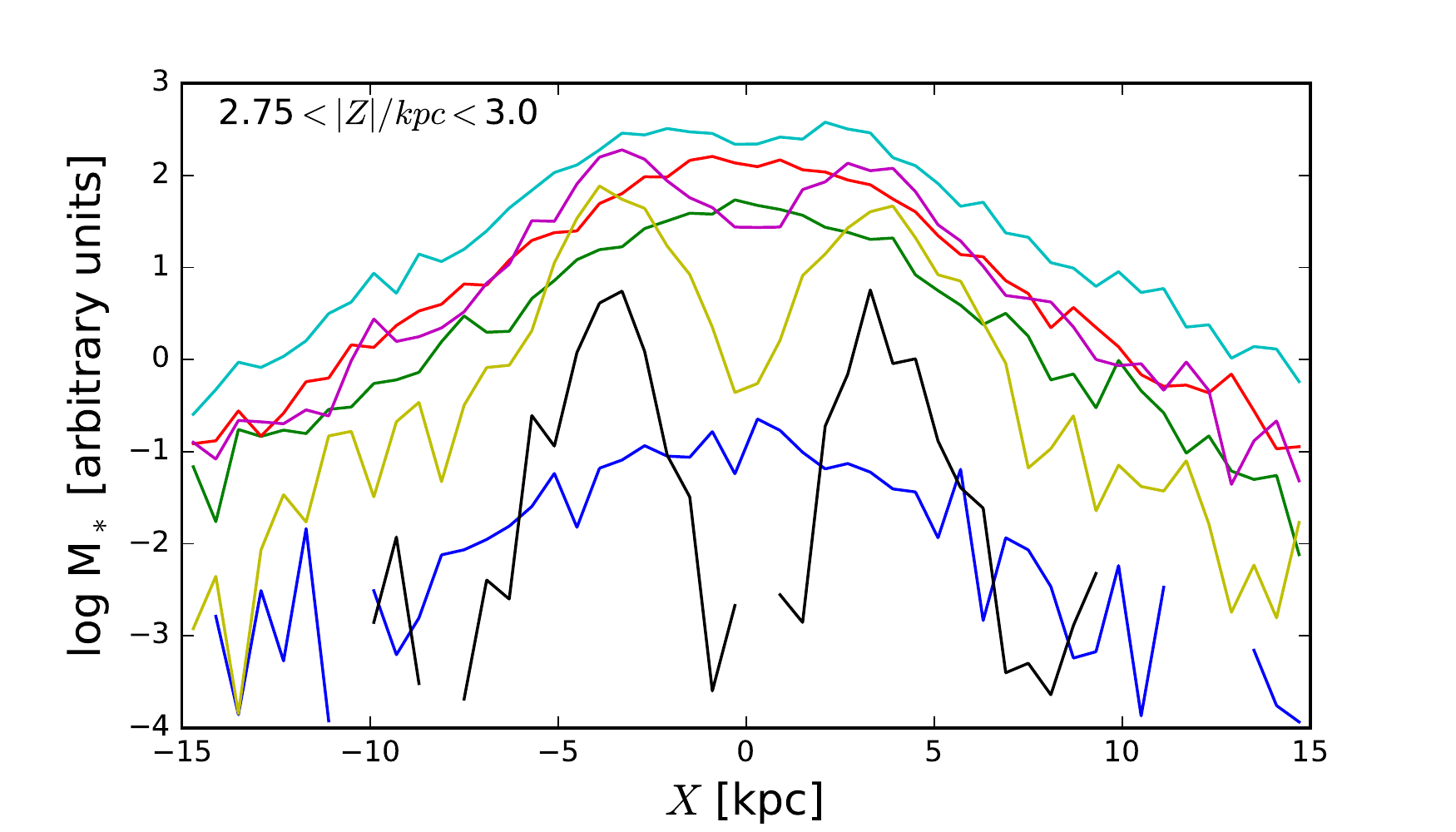}
}
\caption{Density profiles along the $X$-axis (along which the bar is
  aligned) for different heights above the mid-plane, as indicated.
  At each height, the profiles are split by the time of formation,
  \tform, of the stars, as indicated at top left.  The distribution is
  shown at redshift $z=0$.  Only particles at $|Y| < 0.5\kpc$ are
  included.
  \label{fig:profiles}}
\end{figure*}

In the MW, the distance distribution of red clump stars has a single
peak at $|b| \la 5\degrees$ \citep{babusiaux_gilmore05, rattenbury+07,
  cao+13, gonzalez+13}, becoming bimodal at $|b| \ga 5\degrees$
(corresponding to $|Z| \simeq 700 \pc$ on the minor axis)
\citep{mcwilliam_zoccali10, saito+11, wegg_gerhard13, ness+13a}.  This
bimodality is strong in metal-rich stars, but absent in metal-poor
stars \citep{ness+12, uttenthaler+12, rojas-arriagada+14}.
\thesim\ shows a dependence on \feh, and also on age \citep[as
  in][]{debattista+17}.  The present day distributions of star
particles of varying ages at different heights are shown in
Fig. \ref{fig:profiles}.  At $|Z| < 0.5\kpc$ the distributions display
only a single peak within the bulge region.  At $0.5 < |Z|/\kpc <
0.75$ stars younger than $4\Gyr$ ($\tform = 13.8 - $age $> 10\Gyr$)
develop a bimodal distribution, whereas distributions of older stars
remain unimodal. At $1.0 < |Z|/\kpc < 1.25$ stars formed at $8 <
\tform/\Gyr < 10$, which are older than the bar itself, first develop
a flat-topped distribution and, above this region, a bimodal one.  At
$1.75 < |Z|/\kpc < 2.0$, the next age bin ($6 < \tform/\Gyr < 8$)
develops a flat-topped distribution with hints of a bimodality further
from the plane.  Stars of yet older ages never develop a bimodality at
least within the region where the number of particles is large enough
to enable such measurements.  The bimodality of older stars appearing
at larger heights was predicted by \citet{pdimatteo16} and
\citet{fragkoudi+17} from their double disc simulations, but the
failure of the oldest stars to exhibit any bimodality to such large
heights is a new result.

\subsection{Development of an X-shaped metallicity distribution}
\label{ss:xfeh}

\begin{figure*}
\centerline{
\includegraphics[angle=0.,width=0.5\hsize]{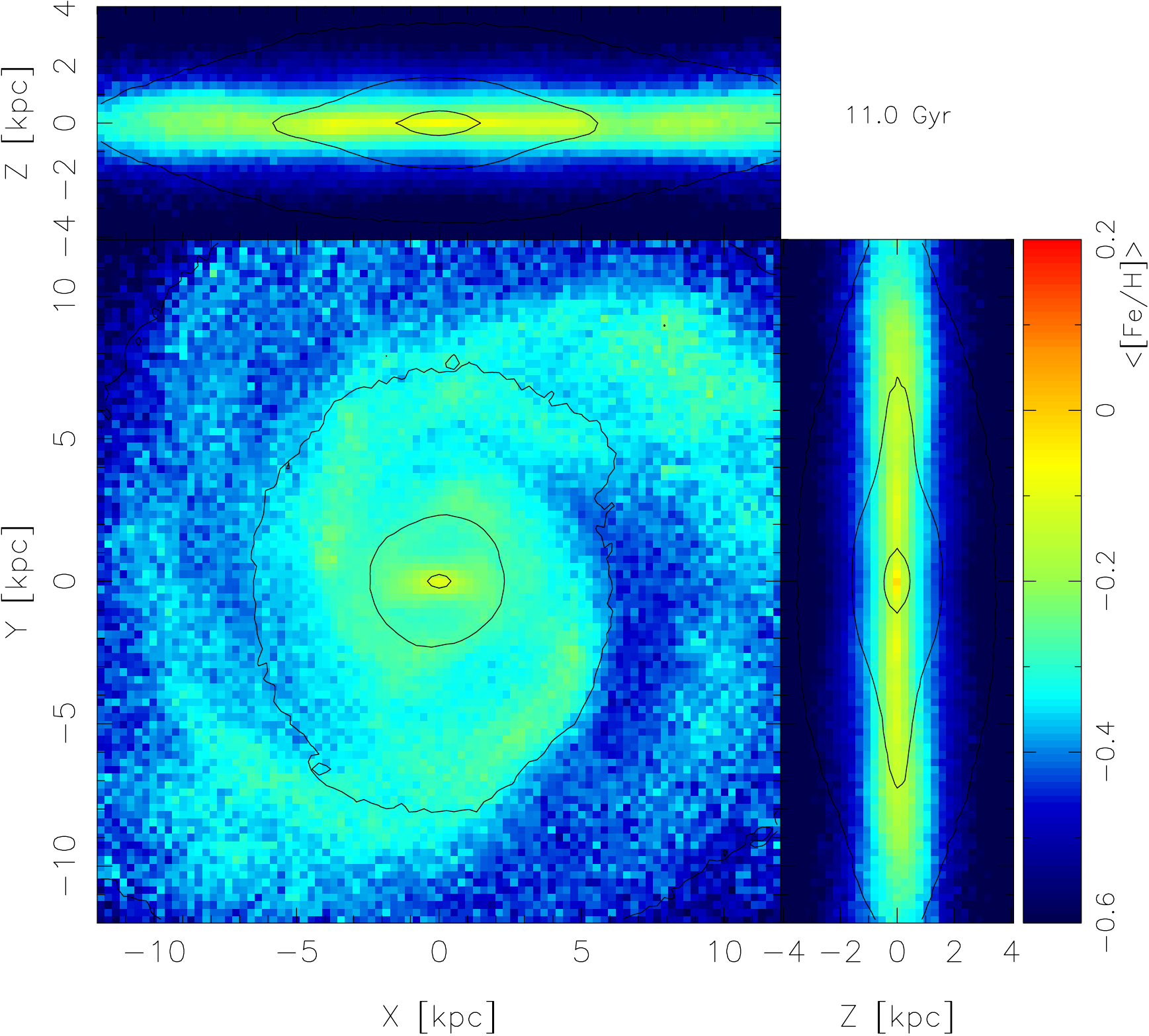}
\includegraphics[angle=0.,width=0.5\hsize]{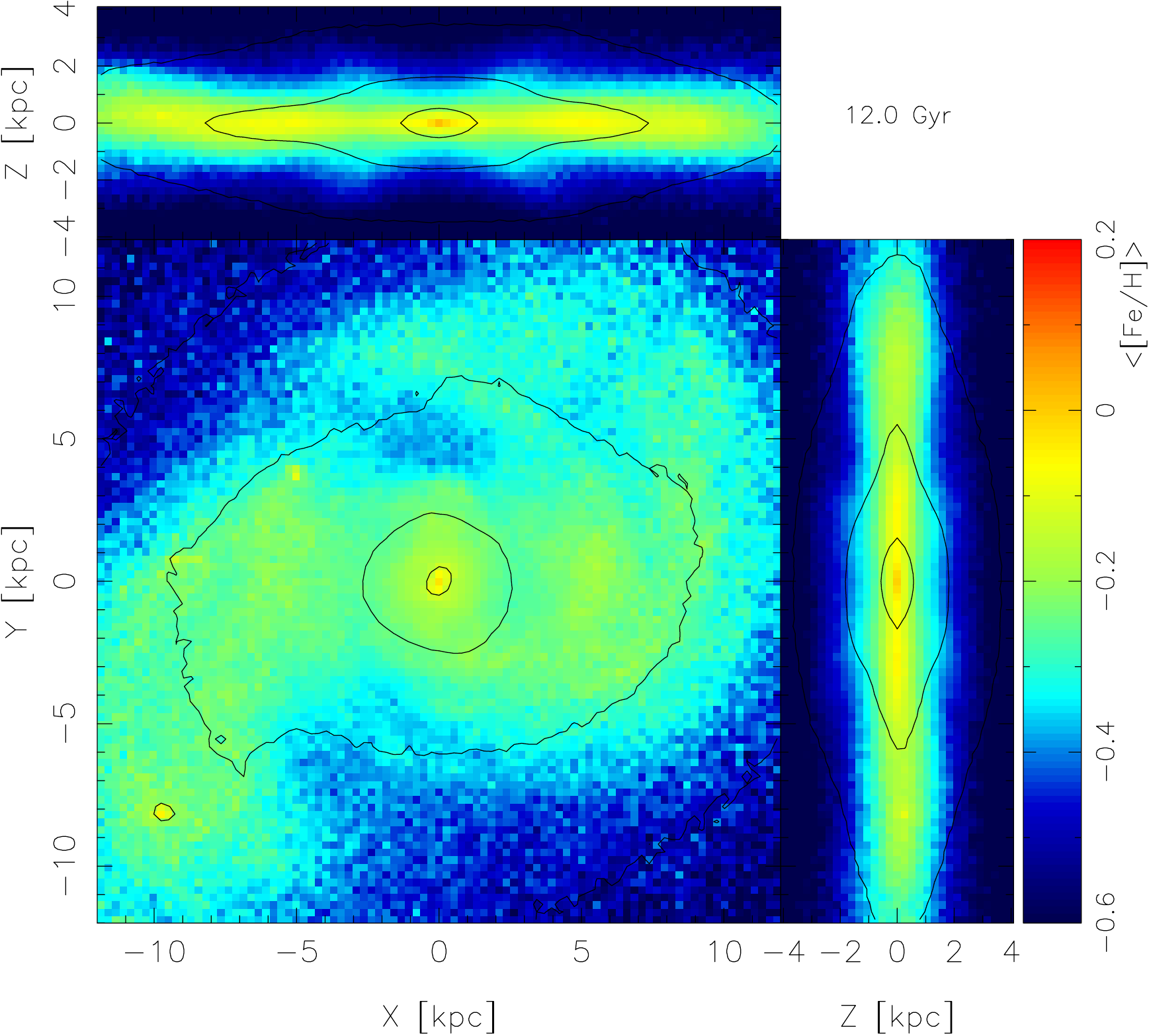}
}
\centerline{
\includegraphics[angle=0.,width=0.5\hsize]{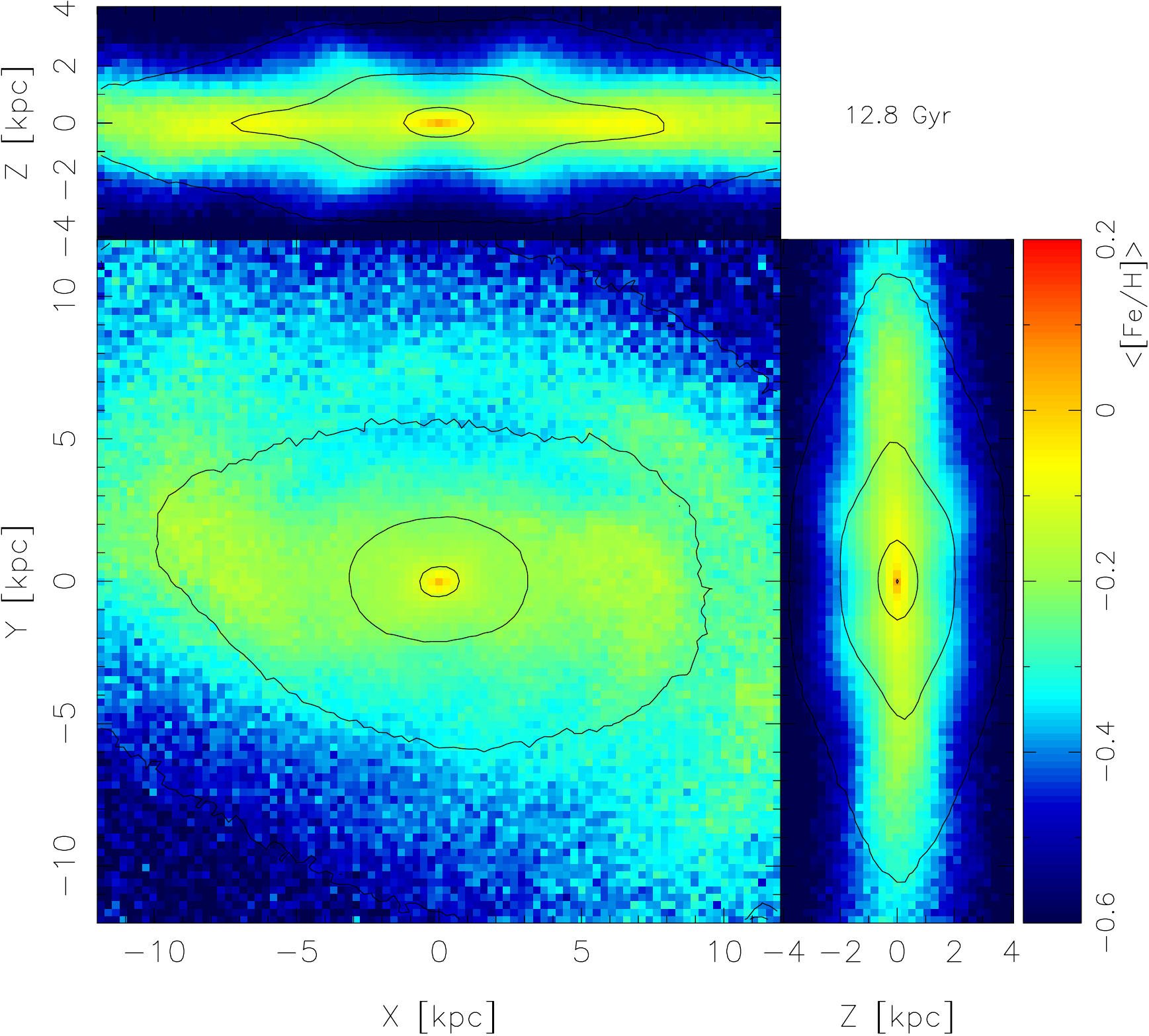}
\includegraphics[angle=0.,width=0.5\hsize]{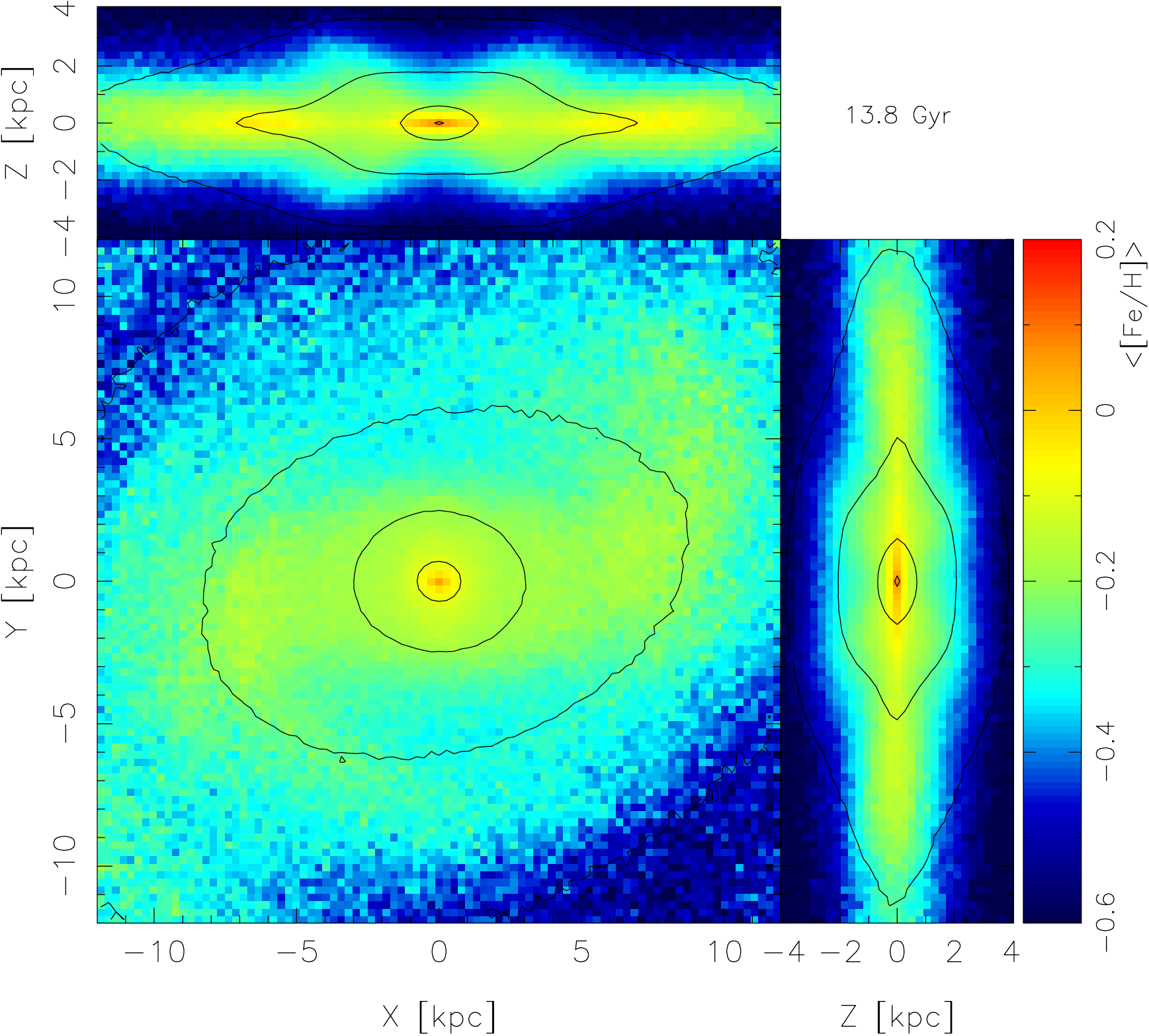}
}
\caption{Orthogonal projections of the metallicity and density, 
spanning the formation of the bar.  The development of the stellar
X-shaped metallicity distribution can be seen in the $(X,Z)$
projection. The projected surface mass densities are indicated by
contours while colours indicate the mean metallicity.  The time,
indicated at top-right in each set of panels, spans from $11 \Gyr$
($z=0.24$) to $13.8\Gyr$ ($z=0$), during which time the bar forms.
\label{fig:metalmaps}}
\end{figure*}

Fig. \ref{fig:metalmaps} shows the evolution from $11\Gyr$ to $13.8
\Gyr$ of the mean metallicity in three orthogonal projections.  At
$11\Gyr$ the bar has not yet developed, and high metallicity stars are
mostly concentrated near the mid-plane, $|Z| < 1\kpc$.  At $12\Gyr$
the bar is forming and an incipient X-shaped \avg{\feh}\ distribution
is evident.  Starting near $X = -5\kpc$ the disc can be seen to be
bending vertically.  Some of this bending continues to $12.8 \Gyr$ by
which point the metallicity distribution has a clear X-shape.  By
$13.8\Gyr$ the bulge is prominently B/P-shaped and an X-shape in the
\avg{\feh}\ distribution is very apparent.  The metallicity
distribution is significantly more peanut-shaped than the density
distribution, an important prediction of kinematic fractionation
\citep{debattista+17} which was confirmed in NGC~4710
\citep{gonzalez+17}.


\section{Vertex Deviation}
\label{s:vertexdeviation}

Because the vertex deviation as a function of metallicity has only
been measured reliably in Baade's Window, at $(l,b) =
(1\degrees,-4\degrees)$, \citep{soto+07, babusiaux+10}, in this
Section we rescale \thesim\ as described in Section
\ref{ss:rescaling}.  After rotating the bar to the Solar perspective,
we select particles in the equivalent of Baade's Window in a
$1\degrees\times 1\degrees$ field.  This field contains $\ga 24,000$
star particles within the distance range $6$ to $10 \kpc$.  Using
these particles, we calculate the vertex deviation, $\theta_v$,
defined as:
\begin{equation}
\tan{2\theta_v}=\frac{2\sigma_{rl}^2}{|\sigma_r^2-\sigma_l^2|}
\label{e:vertdev}
\end{equation}
where $\sigma_{r}^2$ and $\sigma_{l}^2$ are the variances of the
velocities across the radial and longitudinal directions and
$\sigma_{rl}^2$ is the covariance between the two.  The vertex
deviation is the angle of the major axis of the velocity ellipsoid
with the radial direction.

\begin{figure}
\centering
{
\includegraphics[scale=0.55,angle=0,trim=0.0cm 0.0cm 0.0cm 0.0cm, clip=true]{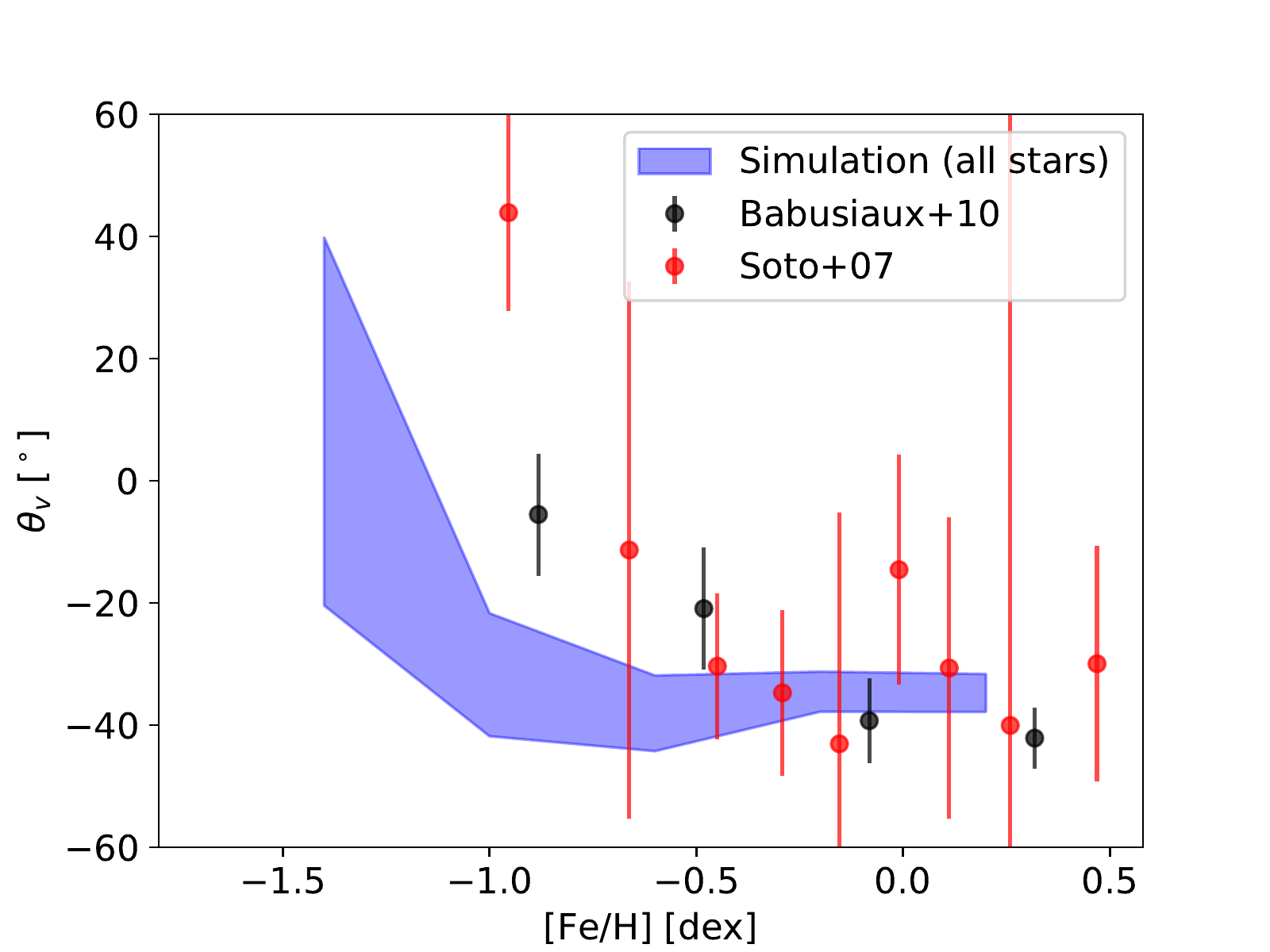} \\
}
\caption{Vertex deviation as a function of metallicity for all stars
  in the rescaled simulation.  Points in black are measurements for
  the MW from \citet{babusiaux+10} and in red from \citet{soto+07}}.
\label{f:vertexfeh}
\end{figure}

Radial velocities and proper motions for star particles are calculated
from their Galactocentric velocities using \textit{galpy}
\citep{galpy}.  As did \citet{babusiaux+10}, we only find an 
insignificant global anisotropy $\sigma_l/\sigma_r \simeq 1.02$,
reaching to $\sim 1.07$ for the youngest stars.  Our value of $C_{lr}
= \sigma^2_{rl}/(\sigma_r \sigma_l)$, where $\sigma_{rl}$ is the
covariance, varies from $\sim -0.25$ for young stars to $\sim 0.02$
for the oldest ones, a decreasing trend found also in the
\citet{babusiaux+10} data.  We obtain the corresponding vertex
deviation using Eq. \ref{e:vertdev}. Fig. \ref{f:vertexfeh} shows
$\theta_v$ as a function of \feh.  While the simulation and the MW do
not match in detail, the general trend of decreasing $|\theta_v|$ for
metal-poor stars is reproduced by the simulation.  At higher
metallicities ($\feh \ga -0.5$ in the MW, and $\feh \ga -1$ in
\thesim), the vertex deviation is roughly constant at $|\theta_v|
\simeq 40\degrees$.  The same trend is also found in the dynamical
model of \citet{portail+17} (their Figure 17).

The vertex deviation $|\theta_v|$ starts declining at a lower \feh\ in
the rescaled simulation compared with the MW.  The metallicity
distribution function in Baade's window in the simulation is similar
to that observed in the MW \citep{zoccali+08}.  However the model's
star formation history, seen in the top panel of Fig
\ref{fig:mdfsfh} is quite different, with star formation peaking
later in \thesim.  This difference in star formation history probably
accounts for the difference in the variation of the vertex deviation
with metallicity.  The constant vertex deviation at the higher
metallicities indicates that the relation between line-of-sight and
longitudinal velocities is unchanged by the strength of the bar, with
only the scatter of the correlation (quantified by $C_{lr}$) varying.

\begin{figure}
\centering
{
\includegraphics[scale=0.55,angle=0,trim=0.0cm 0.0cm 0.0cm 0.0cm, clip=true]{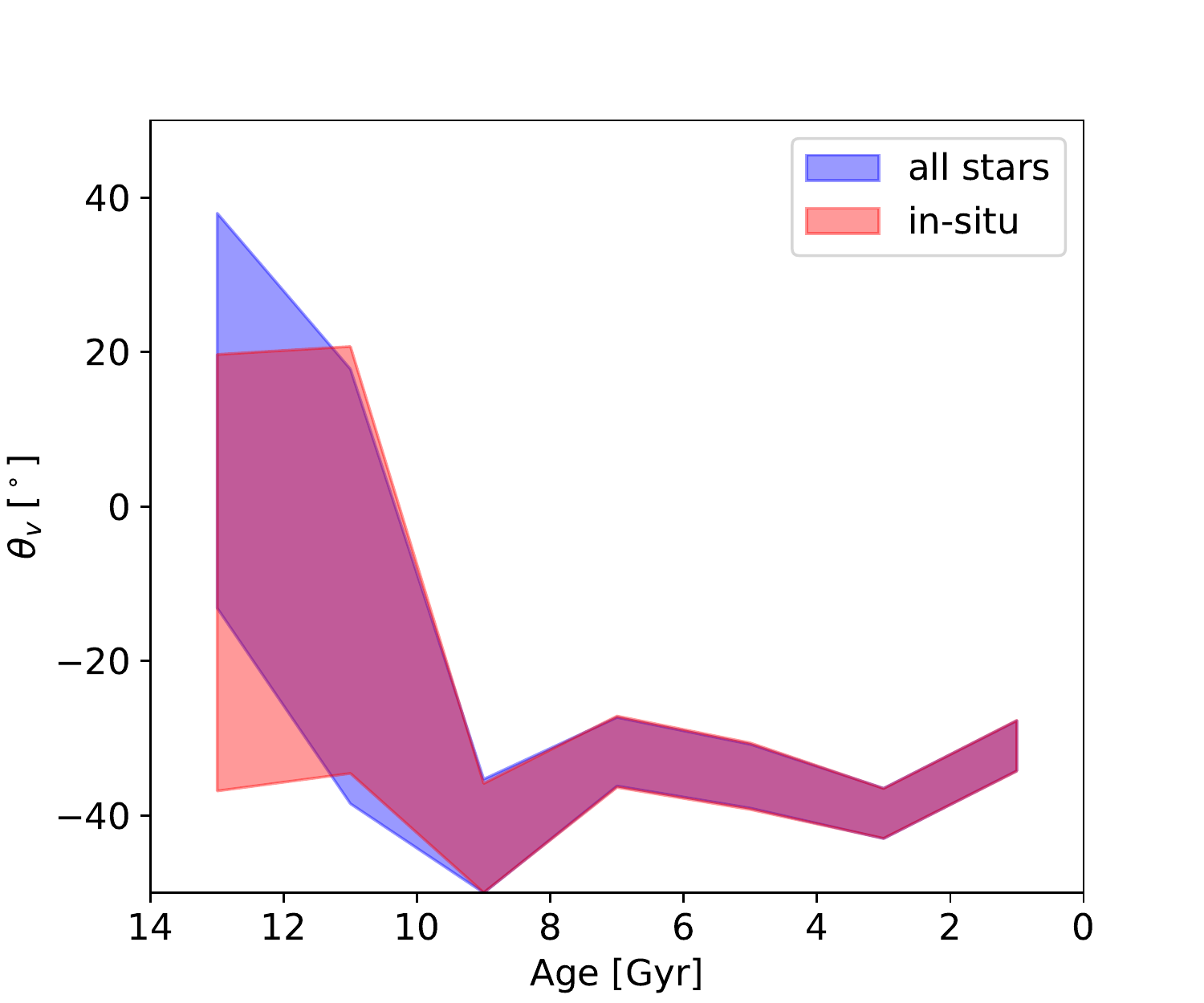}
\includegraphics[scale=0.55,angle=0,trim=0.0cm 0.0cm 0.0cm 0.0cm, clip=true]{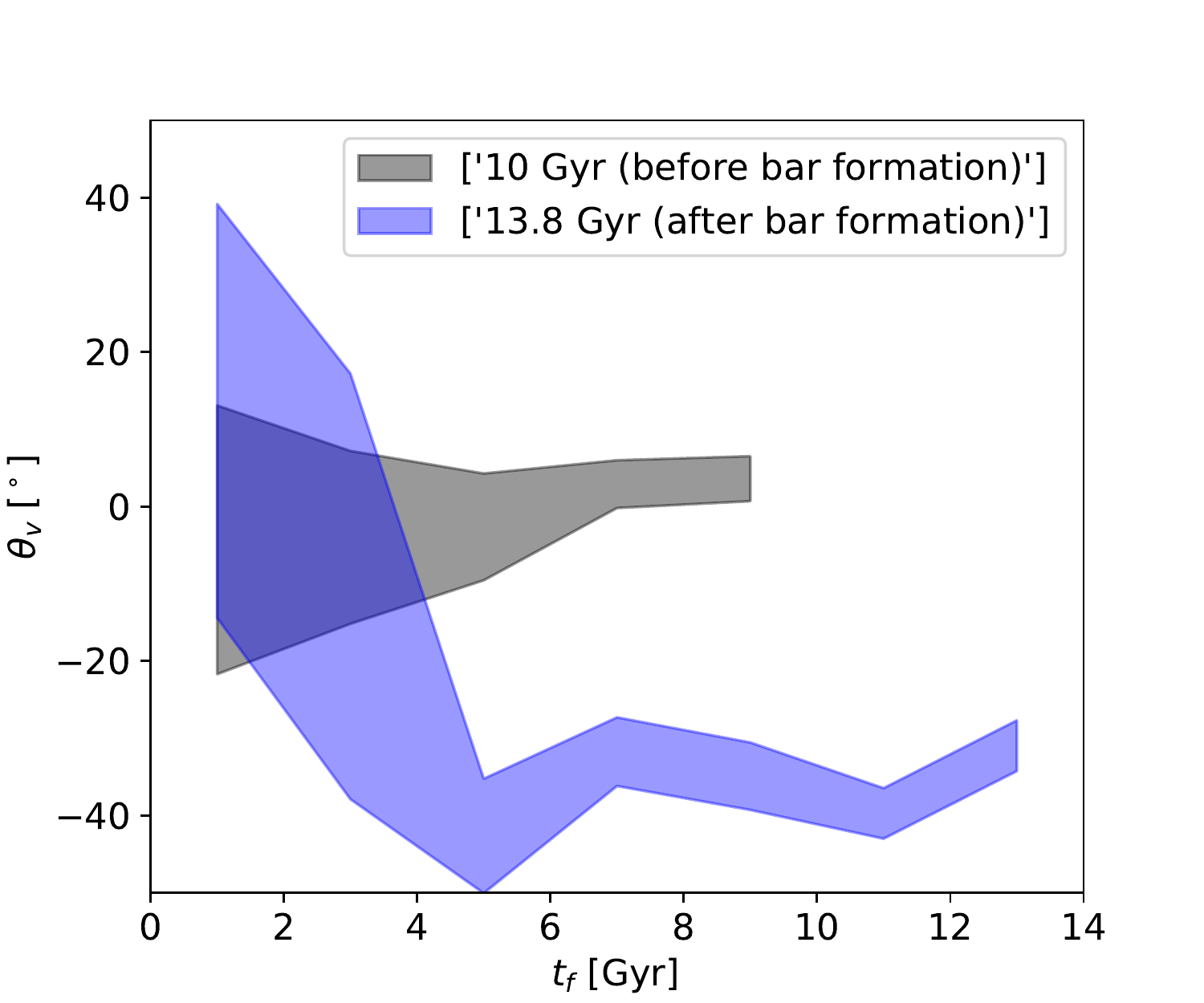}
}
\caption{Top: Vertex deviation in the rescaled simulation as a
  function of age.  The shaded intervals show $\theta_v$ for all stars
  (blue) and stars formed in situ (red).  Bottom: Vertex deviation in
  the rescaled simulation as a function of time of formation of the
  stars.  The blue band is for all stars at $t = 13.8 \Gyr$ ($z = 0$)
  while the grey band is for the simulation before the bar has formed,
  at $t=10\Gyr$ ($z = 0.34$) rescaled using the same factor.}
\label{f:vertexage}
\end{figure}

\citet{debattista+17} showed that many of the trends with metallicity
observed in the MW are fundamentally trends with age, which correlates
with metallicity \citep{bernard+18}.  In the top panel of
Fig. \ref{f:vertexage} we plot the vertex deviation as a function of
stellar age.  The old stars (age $>12 \Gyr$) exhibit a negligible
vertex deviation that increases to $|\theta_v| \sim 40\degrees$ with
decreasing stellar age.  The vertex deviation is large for populations
as old as $9\Gyr$; the bar therefore is comprised of stellar
populations much older than the bar itself.  While the star formation
history peaks at ages $\sim 6\Gyr$, we note that the smallest
uncertainty in $|\theta_v|$ is at the youngest stars.  The uncertainty
therefore represents the scatter in the relation between radial and
longitudinal motion, rather than particle number statistics.
To test whether an accreted old component is responsible for the
observed dependence on age, we measure the vertex deviation for stars
that formed in situ, which we now conservatively define as those stars
formed at Galactocentric distances smaller than $20\kpc$.  Fig.
\ref{f:vertexage} shows that old stars formed in situ show a
negligible difference from the case when all stars are included,
despite the fact that accreted stars are $\sim 60\%$ of all stars
formed before $t_f = 3\Gyr$.  This is similar to the result of
\citet{elbadry+18a} who found the same kinematics for accreted and
in-situ stars of the same age.  Younger stars also show no significant
change when only in-situ stars are chosen, since they dominate at this
age.  This demonstrates, using a fully cosmological simulation, that
the vertex deviation of the velocity ellipsoid of old (metal-poor)
stars in the MW's bulge does not require an accreted bulge component.
Nonetheless, the time at which $\theta_v$ becomes nearly zero is
comparable to the time of the last major merger event.  Our results
therefore do not exclude that it was originally a merger that heated
the bulge to produce the trends observed.  Indeed \thesim\ has the
largest number of satellites of any of the $\sim 15$ FIRE simulated
galaxies in this mass range \citep{garrison-kimmel+18b}.

The bottom panel of Fig. \ref{f:vertexage} shows that before the bar
forms no population exhibits a non-zero $\theta_v$, as is expected for
a stationary, axisymmetric system \citep[e.g.][]{soto+07}.  The
presence of the bar therefore drives the vertex deviation; the small
$\theta_v$ in the oldest stars is just a consequence of the weak bar
in this population, as seen in Fig. \ref{fig:densitybyage}.


\section{Constraint on the Age of the Milky Way's bar}
\label{s:agemwbar}

\begin{figure}
\centerline{
\includegraphics[angle=0.,width=\hsize]{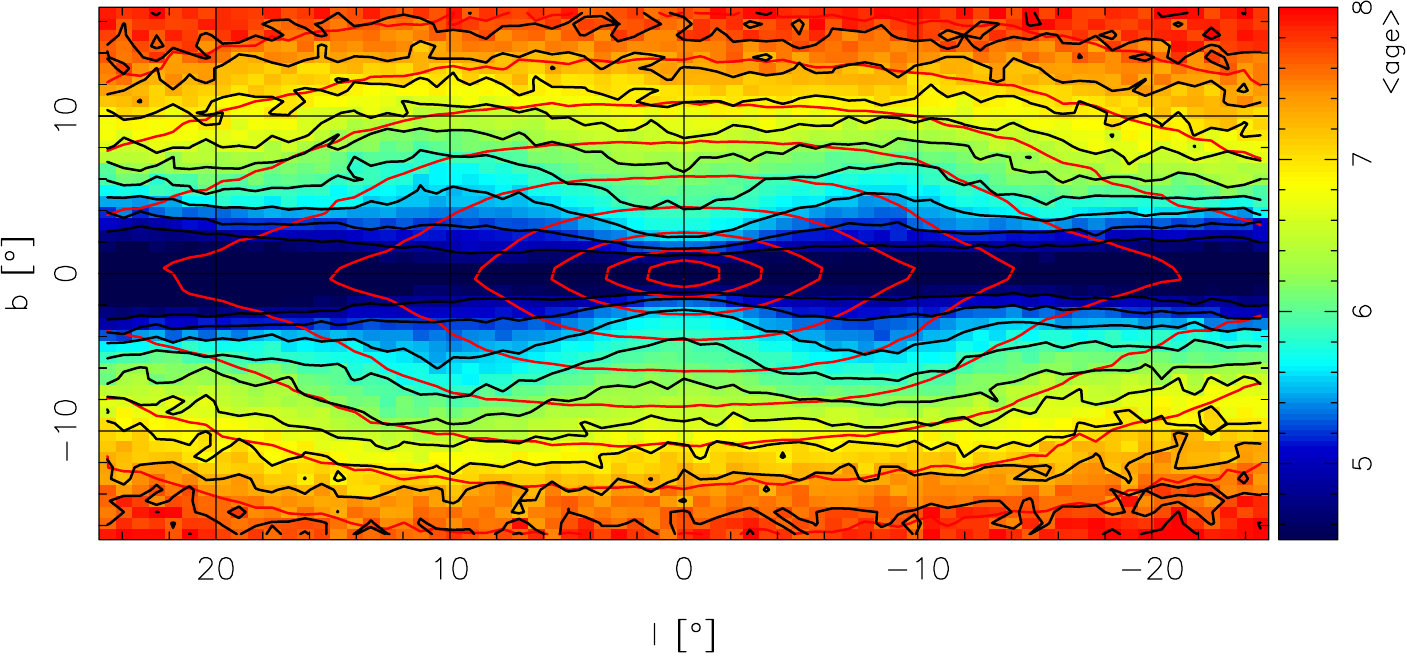}
}
\centerline{
\includegraphics[angle=0.,width=\hsize]{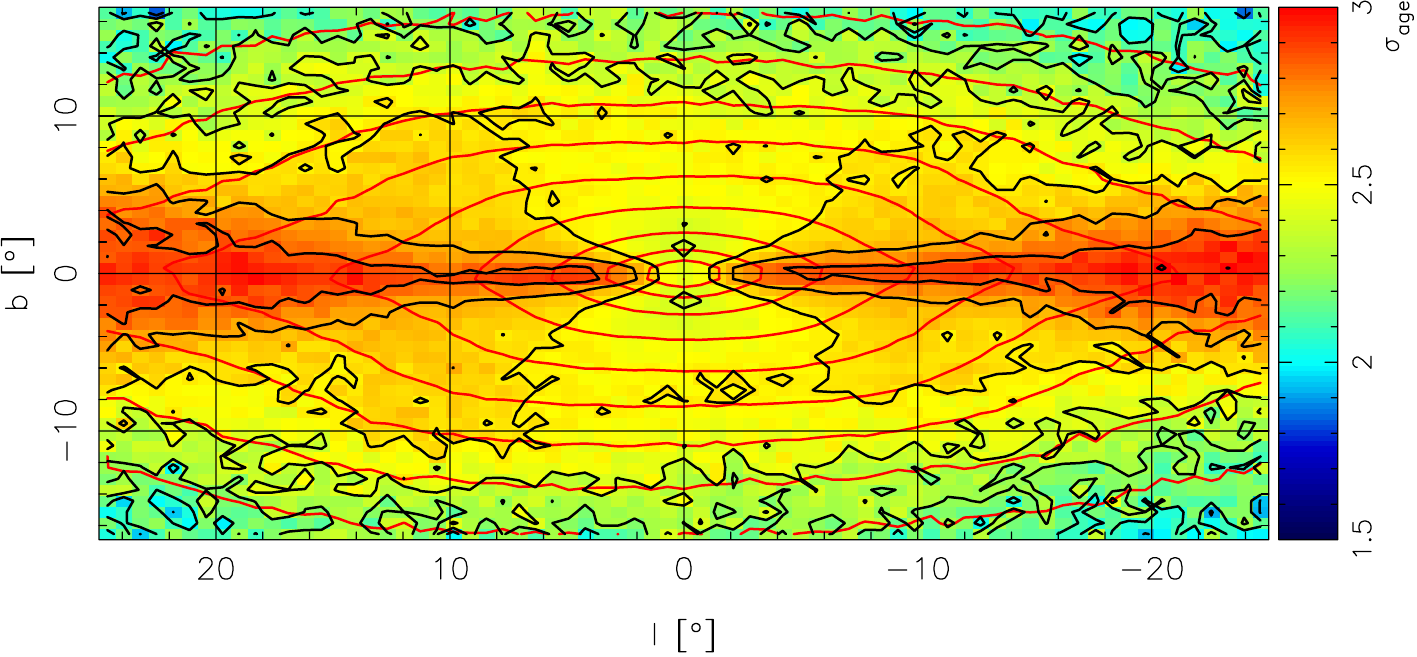}
}
\caption{Maps in $(l,b)$ space of the mean age (top) and age
  dispersion (bottom) at $z=0$, with the model scaled as described in
  Section \ref{ss:rescaling} and oriented to the Sun's viewing angle.
  Red contours show the surface density of the model while the black
  contours are for the plotted quantity.
  \label{fig:agemoments}}
\end{figure}

\citet{sheth+08} found that the barred fraction amongst high mass
galaxies has barely changed since redshift $z \sim 0.84$, \ie\ $\sim
7\Gyr$ ago \citep[see also][]{erwin18}.  It is important therefore to
explore whether the MW's bar is also this old.  Here we show that the
ability of the bar to thicken pre-existing populations means that the
bar in the MW cannot be young.

Fig. \ref{fig:agemoments} shows the mean age and age dispersion of the
model at $z=0$ with the model scaled and oriented to the MW.  The mean
age at large heights, $|b| \ga 10\degrees$, is $\ga 7\Gyr$ and
decreases slowly to larger heights.  Meanwhile the age dispersion is
$\sim 2 - 2.5 \Gyr$ at these heights.  The simulation of
\citet{debattista+17}, which formed a bar much earlier in its history,
has a comparable mean age at these heights.  However the typical age
dispersion is lower, $1\Gyr$.  This suggests that a significant tail
of young stars will be found at these large heights in \thesim.

\begin{figure}
\centerline{
\includegraphics[angle=0.,width=\hsize]{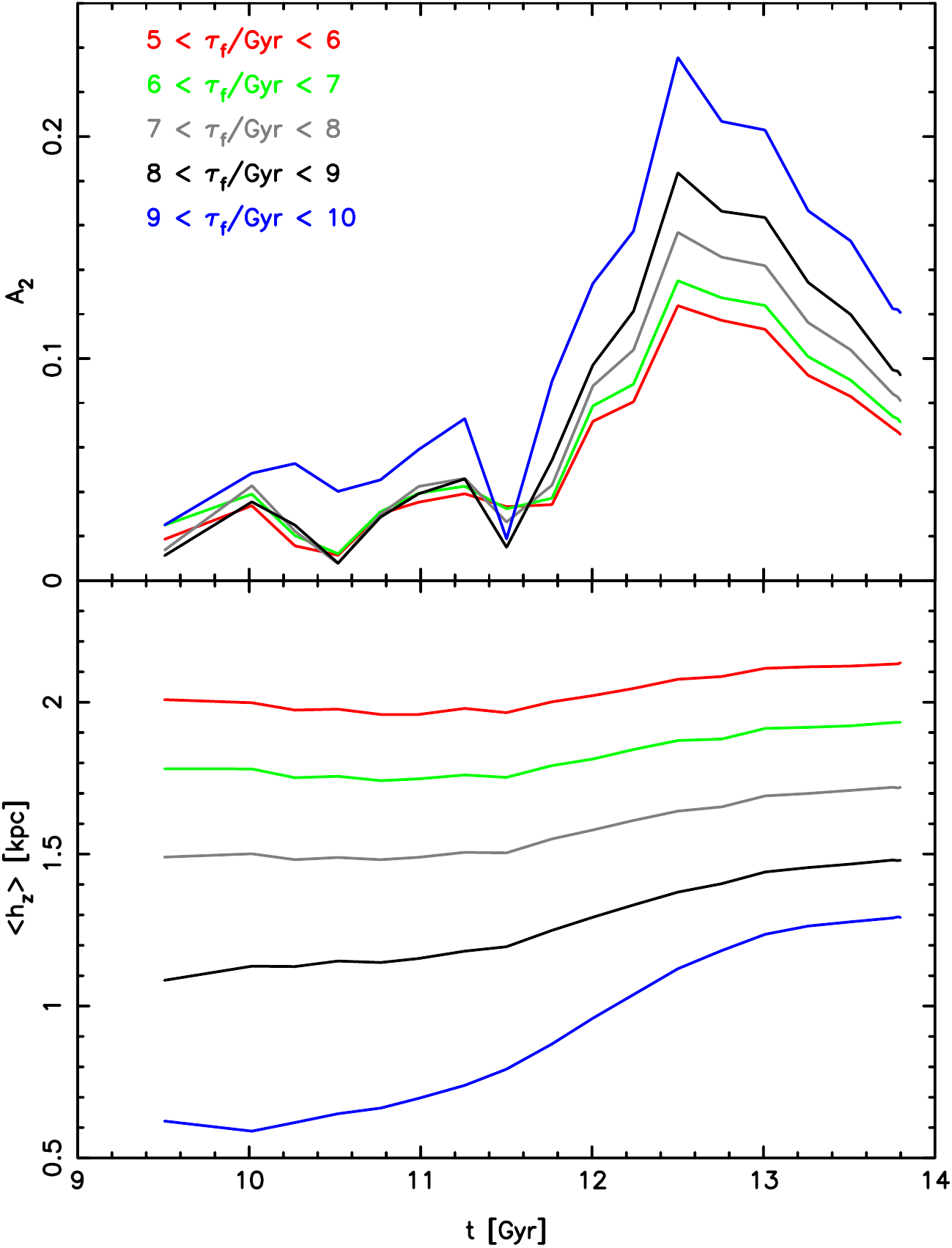}
}
\caption{The evolution of bar amplitude (top) and average
  root-mean-square height (bottom) of stars.  Stars are separated by
  time of formation, $\tform = 13.8\Gyr-$age.  The younger stars form the
  strongest bar.  All populations are vertically heated by the bar.
  \label{fig:fractionation}}
\end{figure}

Fig. \ref{fig:fractionation} shows the evolution of stellar
populations that formed between $5\Gyr$ and $10\Gyr$.  The younger
populations form the strongest bar, as seen in the density maps of
Fig. \ref{fig:densitybyage} and Fig. \ref{fig:baramps}.
\citet{debattista+17} attribute this behaviour to the lower radial
velocity dispersion of the younger stars at the time of bar formation.
The evolution of the average heights, $\avg{h_z}$, averaged in the
radial range $1 < R/\kpc < 6$, is shown in the bottom panel.  The
young populations are thinner, as expected \citep[see also][]{ma+17}.
The onset of bar formation between $11\Gyr$ and $12\Gyr$ leads to a
steepening of the vertical heating of all the populations, but is most
prominent for the young populations.  Nonetheless, younger populations
remain thinner, as required by kinematic fractionation.

\begin{figure*}
\centerline{
\includegraphics[angle=0.,width=0.5\hsize]{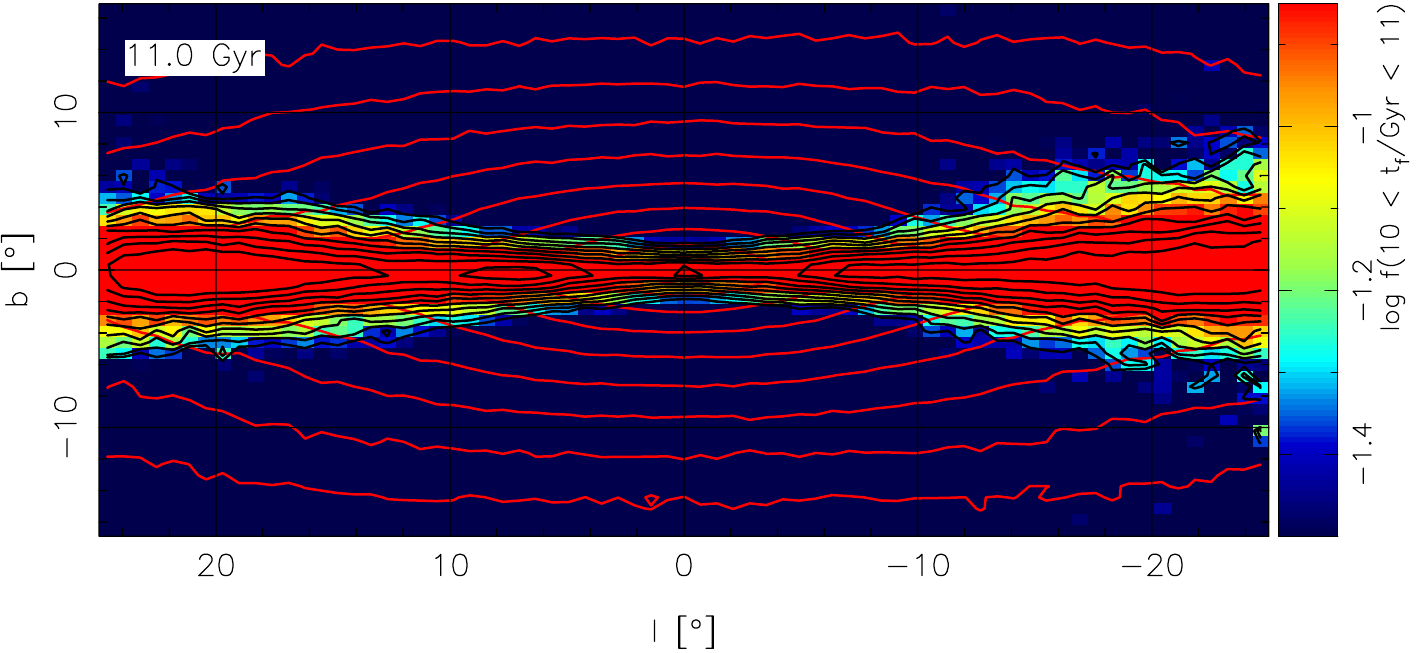}
\includegraphics[angle=0.,width=0.5\hsize]{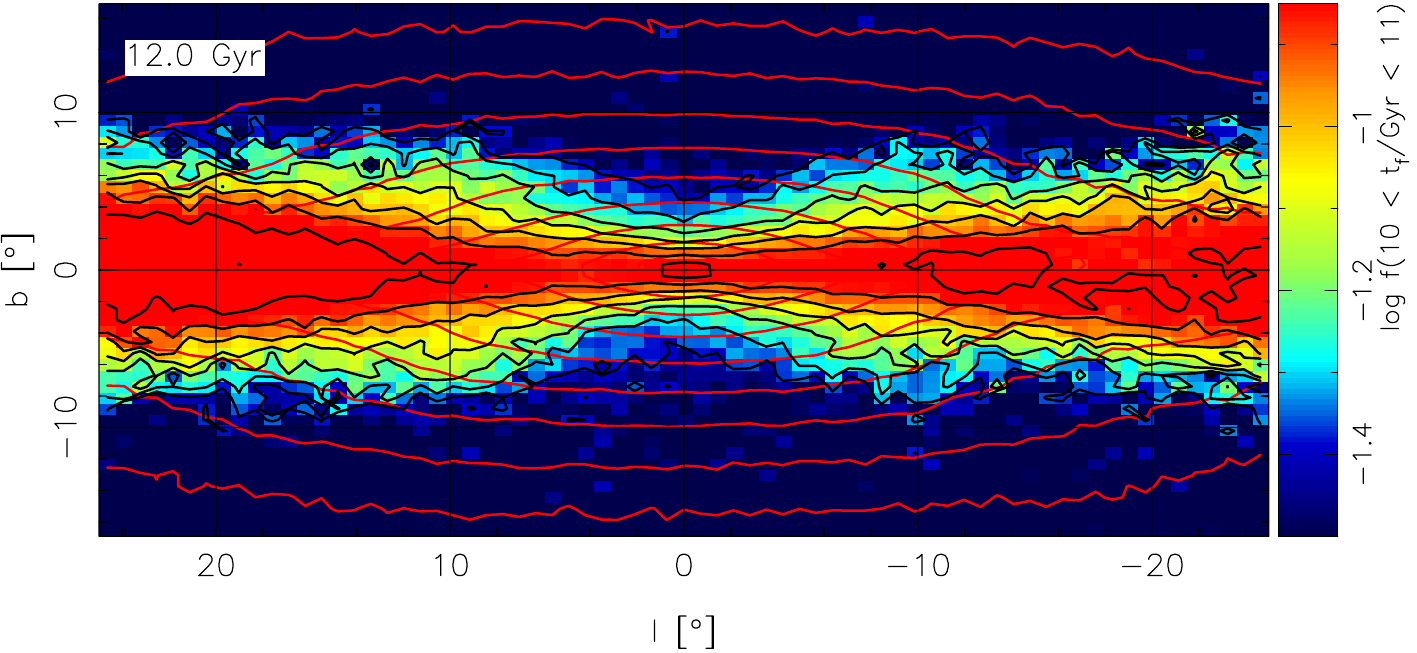}
}
\centerline{
\includegraphics[angle=0.,width=0.5\hsize]{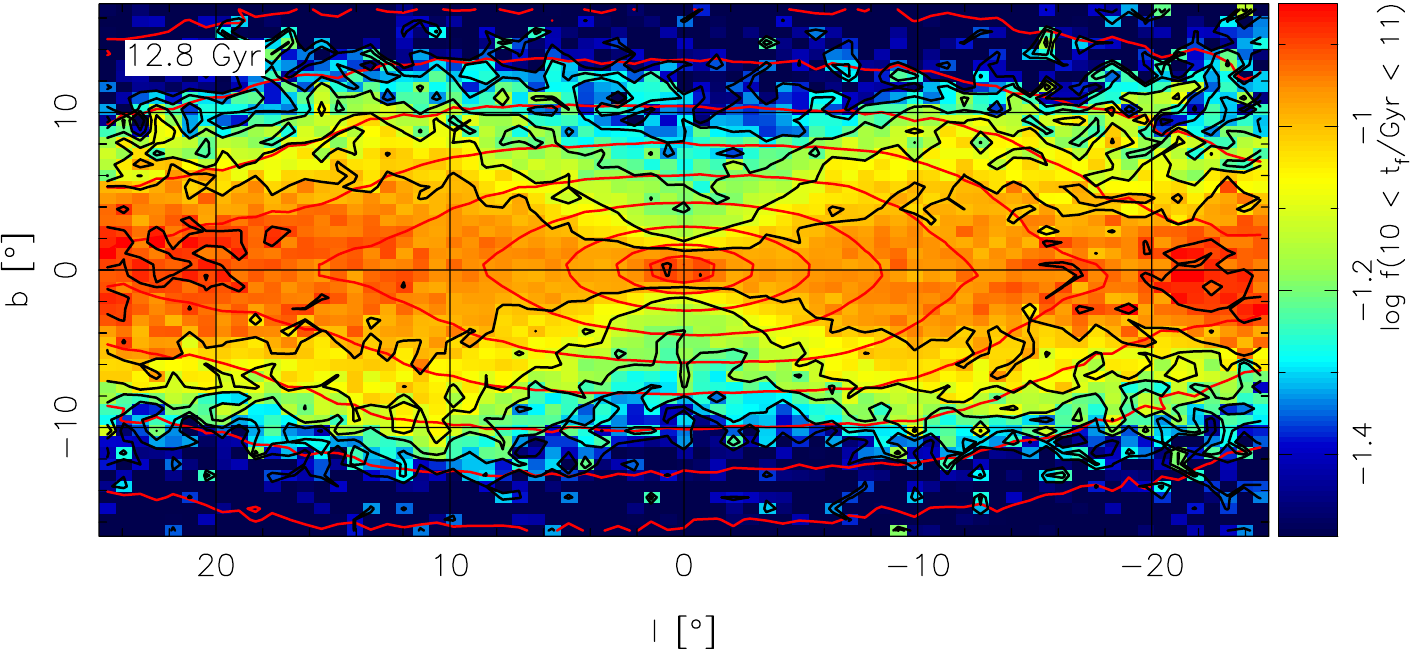}
\includegraphics[angle=0.,width=0.5\hsize]{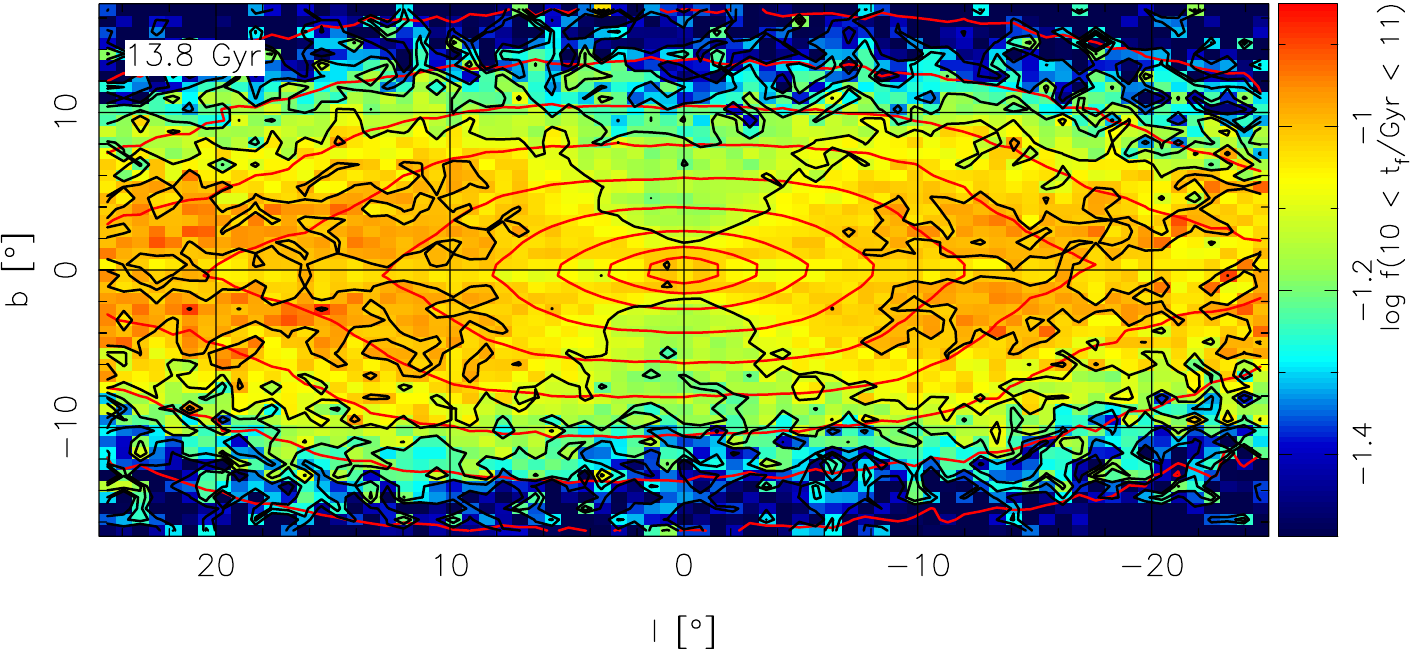}
}
\caption{Maps of the fraction of stars born in the time interval $10
  \leq \tform/\Gyr \leq 11$; the model has been scaled identically at
  each timestep to approximate the MW's X-shape at $z=0$, as described
  in Section \ref{ss:rescaling}.  The snapshots are at $11\Gyr$ (top
  left) to $13.8 \Gyr$ (bottom right).  Red contours indicate the
  surface density, as seen from the Solar orientation, which is
  identical in all panels.
  \label{fig:fintermediate}}
\end{figure*}

The strong vertical heating by the bar dredges relatively young stars
into the line of sight of the bulge.  With \thesim\ scaled as
described in Section \ref{ss:rescaling}, and the bar oriented at
$27\degrees$ to the line of sight to the Galactic centre
\citep{wegg_gerhard13}, we map in Fig. \ref{fig:fintermediate} the
evolution of the fraction of stars that formed during the time
interval $10 \leq t_\mathrm{f}/\Gyr \leq 11$ across the bulge, 
\ie\ shortly before the bar starts forming.  While a negligible
fraction of stars this age are found on the minor axis shortly after
they form, as the bar strengthens their fraction grows rapidly.  Such
a fraction of stars that are only $2.8 - 3.8\Gyr$ old now would be
obvious, {\it particularly at $|b| > 8\degrees$}, if it were present
in the MW.

\begin{figure}
\centerline{
\includegraphics[angle=0.,width=\hsize]{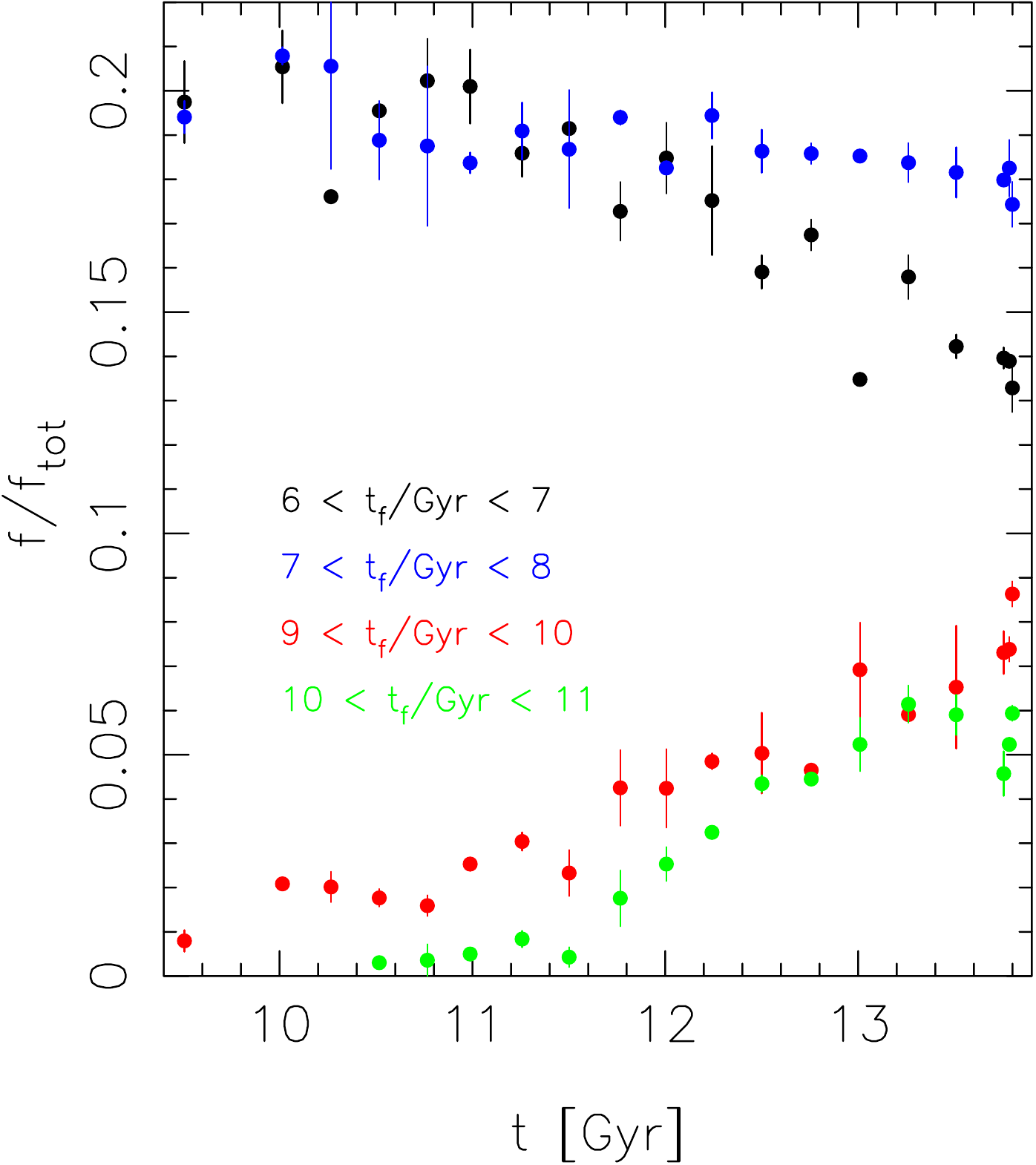}
}
\caption{The effect of the bar on the evolution of the fraction of the
  stellar populations at $|b| = 10\degrees$ on the minor axis in the
  rescaled version of \thesim.  Stars are separated by time of
  formation, $\tform = 13.8\Gyr-$age.  Each measurement is within a
  window of $42\arcsec \times 42 \arcsec$.  Error bars are based on
  the difference between $b = +10\degrees$ and $b = -10\degrees$.  The
  bar starts forming at $11.5\Gyr$ (see Fig. \ref{fig:barformation}).
  \label{fig:vheating}}
\end{figure}

Fig. \ref{fig:vheating} shows the evolution of the fraction of stars
of various ages on the minor axis at $|b| = 10\degrees$.  The fraction
of stars that form after $9\Gyr$ (which are $4.8\Gyr$ or younger at
present) rises sharply after the bar starts forming.  Overall the
fraction of stars born after $t=10 \Gyr$ ($z = 0.34$) reaches $\sim
15\%$, considerably more than previously suggested in the MW
\citep[e.g.][]{ortolani+95, kuijken_rich02, zoccali+03, sahu+06,
  clarkson+08, clarkson+11, brown+10, valenti+13, calamida+14}.  Such
a fraction of young stars in the bulge has indeed been suggested by
recent measurements \citep{bensby+17, haywood+16, bernard+18}, but
only at low Galactic latitudes ($\rm |b|<4^\circ$) and at high
metallicities. In particular, the age-metallicity relation presented
in \citet{bernard+18} \citep[which is consistent with the microlensed
  dwarfs from][]{bensby+17} shows that young stars are also those in
the near-Solar metallicity range ($\rm -0.2<[Fe/H]<0.5$). This is the
dominant population at low latitudes, where the fraction of young
stars ($\sim15\%$) is observed \citep{haywood+16}, but it weakens with
increasing Galactic latitude, and is marginal at $|b|=10\degrees$
\citep{ness+13a, zoccali+17}.  A late-forming bar therefore
excessively contaminates the bulge with relatively young stars to high
latitudes.  Stars that form at $10\Gyr$ are only $1.5 \Gyr$ old by the
time the bar starts forming in \thesim; they are therefore unlikely to
have been vertically heated excessively by either physical or
numerical effects.  The number of them that reach large height
therefore is probably a quite robust result that does not depend
strongly on the details of the model's evolution once it is scaled to
the size of the MW.  Indeed in the model of \citet{debattista+17},
stars forming before the bar are a major component of the bulge at
large height.  In \thesim, the star formation drops significantly
after $9\Gyr$, as seen in Fig. \ref{fig:mdfsfh}; nonethess $23\%$ of
stars are younger than $3.5 \Gyr$ ($t_f > 10 \Gyr$). If this were a
factor of $\sim 2$ lower \citep[e.g.][]{snaith+15}, the fraction of
stars at $|b| = 10\degrees$ would still be too high compared to the
MW.  We conclude that the MW's bar could not have formed as recently
as in \thesim\ if the bulge lacks a young population at high
latitudes.


\section{Discussion and Conclusions}
\label{s:discussion}

A non-zero vertex deviation $\theta_v$ arises once the bar forms.  At
the low metallicity end the variation in $\theta_v$ with age (and
metallicity) is due to the difference in bar strength that results
from populations with different random motions at the time of bar
formation.  It is in this sense another manifestation of kinematic
fractionation, the separation of stellar populations on the basis of
their kinematics, rather than being a signature of an accreted
population in the bulge.  However an accreted population settles into
a hot component and would therefore also produce the same signature,
so our results do not exclude an external origin for the zero vertex
deviation component in the Milky Way.  At the high metallicity end,
the vertex deviation is constant, in \thesim\ as in the MW, because a
strong bar is present.  The increasing bar strength with metallicity
at this end of the relation is reflected in the decreasing uncertainty
on the vertex deviation.

The maximum vertex deviation in \thesim\ and the Milky Way are
comparable, $|\theta_v| \sim 40\degrees$.  However $\theta_v$ in the
Milky Way starts decreasing at a larger \feh\ than in the model.
Since a large $|\theta_v|$ is possible only if the bar is strong in a
particular population, the bar must be strong to lower metallicities
in the model than in the Milky Way.  In the Milky Way the population
of stars at $\feh \simeq -1$ is dominated by the stellar halo
\citep{ness+13a}.  The metallicity distribution function of the
rescaled \thesim\ at Baade's Window is not much different from that in
the Milky Way (Fig. \ref{fig:mdfsfh}).  However the star formation
peaks at $\sim 8\Gyr$, which probably accounts for the $|\theta_v|$
turnoff at lower metallicity in \thesim.  In this sense the vertex
deviation may be a quite sensitive probe of the chemical enrichment
and dynamical history of the inner disc before the bar formed.  This
would require measurement of the vertex deviation across a broader
part of the bulge to help understand the strength of the bar better.

The long-held view that the bulge is comprised of only old ($\sim
10\Gyr$ old) stars has recently been challenged, starting with the
discovery of young to intermediate-age stars in microlensing surveys
\citep{bensby+11, bensby+13, bensby+17}.  In their simulation,
\citet{debattista+17} showed that the age distribution of stars in the
bulge is dominated by old stars, with the fraction of stars between
$1$ and $4\Gyr$ old less than $10\%$ everywhere above $|b|\simeq
5\degrees$, while the young stars are concentrated towards the
mid-plane \citep{ness+14}.  In comparison, \thesim, the simulation
studied in this paper, has $\sim15\%$ of young to intermediate-age
stars all the way up at $|b| \sim 10\degrees$.  Despite the large
differences between the bar formation and star formation histories of
these two simulations (and presumably also the MW), the qualitative
similarities in their stellar populations on their minor axes provide
important information on the time when the bar formed.  Indeed the
results here and in \citet{debattista+17} show that stars formed
before and during bar formation are efficiently transported to large
heights and are therefore likely to be found on the minor axis in
significant numbers.  A comparison of Fig. \ref{fig:fintermediate}
here and Fig. 22 of \citet{debattista+17} reveals that a particularly
fruitful place to search for younger populations is at $l\sim
10\degrees$, which is most contaminated by them in both simulations;
this roughly corresponds to the location of the end of the X-shape on
the near-side of the bar.  This region has the further benefit that
obscuration is significantly less severe.  A useful strategy would be
to compare the age distribution, at fixed latitude, at $l \sim
10\degrees$ and on the minor axis, which results in a relatively large
contrast in the fraction of the younger populations.

\subsection{Summary}

Our results can be summarised as follows:
\begin{itemize}
\item We confirm the trends produced by kinematic fractionation. Both 
  the bar strength and the distance bimodality (X-shape) decrease in
  strength with stellar age.  Observed edge-on with the bar side-on
  the metallicity distribution is more peanut-shaped than the density
  distribution itself, as observed in NGC~4710.  We find that
  kinematic fractionation occurs in a fully cosmological context (see
  Section \ref{s:kfractionation}) and must therefore have occurred in
  the Milky Way.
\item We find that a non-zero vertex deviation of the velocity
  ellipsoid at the location of Baade's Window develops when the bar
  forms.  The vertex deviation varies with metallicity, reaching zero
  for metal-poor stars, as in the Milky Way.  The vertex deviation is
  a function of age, reaching $\sim 30-40\degrees$ for stars younger
  than $10\Gyr$, but vanishing for stars older than $10\Gyr$.  As in
  the MW, the vertex deviation is roughly constant for metal-rich
  stars ($\feh \ga -1$ in \thesim, and $\feh \ga -0.5$ in the MW).
  The vanishing vertex deviation of metal-poor stars is not due to an
  accreted population of stars, but to the weak bar in the oldest
  stars, and is also a result of kinematic fractionation (see Section
  \ref{s:vertexdeviation}).
\item A bar forming after redshift $z=0.2$ drives a large fraction of
  stars younger than $4.8 \Gyr$ to large heights on the minor axis of
  the bulge.  Since the fraction of such stars in the Milky Way is
  negligible at high latitudes, we conclude that its bar is very
  likely to have formed before this time.  The Milky Way's bar
  therefore {\it cannot} be young (see Section \ref{s:agemwbar}).
\end{itemize}

\bigskip
\noindent
{\bf Acknowledgements.}

\noindent
V.P.D. was supported by STFC Consolidated grant ST/M000877/1.
R.E.S. was supported by an NSF Astronomy and Astrophysics Postdoctoral
Fellowship under grant AST-1400989.
Support for S.G.K. was provided by NASA through Einstein Postdoctoral
Fellowship grant PF5-160136 awarded by the Chandra X-ray Center, which
is operated by the Smithsonian Astrophysical Observatory for NASA
under contract NAS8-03060.
A.W. was supported by NASA through grants HST-GO-14734 and
HST-AR-15057 from STScI.
Support for P.F.H. was provided by an Alfred P. Sloan Research
Fellowship, NSF Collaborative Research Grant \#1715847 and CAREER
grant \#1455342.
K.E.B. acknowledges support from a Berkeley graduate fellowship, a
Hellman award for graduate study, and an NSF Graduate Research
Fellowship.
Numerical calculations were run on the Caltech compute cluster
``Wheeler,'' allocations from XSEDE TG-AST130039 and PRAC NSF.1713353
supported by the NSF, NASA HEC SMD-16-7592, and the High Performance
Computing at Los Alamos National Labs.
CAFG was supported by NSF through grants AST-1412836, AST-1517491,
AST-1715216, and CAREER award AST-1652522, by NASA through grant
NNX15AB22G, and by a Cottrell Scholar Award from the Research
Corporation for Science Advancement.


\bigskip 
\noindent

\bibliographystyle{aj}
\bibliography{ms.bbl}

\label{lastpage}

\end{document}